\documentclass[11pt]{article}
\usepackage{amssymb}
\usepackage{amsmath}
\usepackage{amsfonts}
\usepackage{mathrsfs}
\usepackage{textcomp}
\usepackage{amsthm}
\usepackage{bookmark}
\usepackage{appendix}
\usepackage{dsfont}
\usepackage{changes}%\usepackage[final]{changes}  %\listofchanges
\usepackage{xcolor}
\usepackage{verbatim}
\usepackage{enumitem}
\usepackage{url}
\usepackage{hyperref}
\usepackage{cases}
\usepackage[T1]{fontenc}
\numberwithin{equation}{section}
\newtheorem{theorem}{Theorem}[section]
\newtheorem{proposition}[theorem]{Proposition}
\newtheorem{lemma}[theorem]{Lemma}

\theoremstyle{remark}

\makeatletter
\renewenvironment{proof}[1][\proofname]{\par
\pushQED{\qed}%
\normalfont\topsep6\p@\@plus2\p@\@minus2\p@\relax
\trivlist
\item[\hskip\labelsep\normalfont\bfseries #1\@addpunct{.}]\ignorespaces
}{%
\popQED\endtrivlist\@endpefalse
}
\makeatother
\newcommand{\R}{\mathbb R}
\newcommand{\Sph}{\mathbb S^2}
\newcommand{\ind}{\mathds{1}}
\newcommand{\TF}{\mathrm{TF}}
\newcommand{\Lip}{\operatorname{Lip}}
\newcommand{\coul}{|\cdot|^{-1}}
\title{Uniform Ionization Bounds in Molecular Thomas--Fermi--Dirac--von Weizs\"acker Theory}
\author{Shuang Wu\thanks{School of Mathematics and Statistics and Hubei Key Laboratory of Mathematical Sciences, Central China Normal University, Wuhan 430079, P. R. China. Email: swu@mails.ccnu.edu.cn.}}
\date{}
\begin{document}
\maketitle

\begin{abstract}
We study the molecular Thomas--Fermi--Dirac--von Weizs\"acker theory. For each fixed number $K$ of nuclei, we prove that every minimizer of mass $N$ satisfies $N\leq Z+C_K$, where $Z$ is the total nuclear charge, and $C_K>0$ is independent of the nuclear charges and positions. This answers the molecular large-charge ionization question posed in [R. L. Frank, P. T. Nam and H. Van Den Bosch, CPAM, 2018]. For $N\geq Z$, we also compare the screened potentials and electronic radii of Thomas--Fermi--Dirac--von Weizs\"acker minimizers with those of the corresponding neutral molecular Thomas--Fermi minimizers and establish a uniform bound on the ionization energy.
\end{abstract}
\medskip
\noindent\textbf{Mathematics Subject Classification (2020).}  81V55, 35Q40, 49J10, 49S05.

\smallskip
\noindent\textbf{Keywords.} {Thomas--Fermi--Dirac--von Weizs\"acker theory, molecular ionization, screened potential, electronic radius, ionization energy.}

\section{Introduction and Main Results}

The ionization conjecture asks whether the number $N>0$ of electrons bound by an atom of nuclear charge $Z>0$ satisfies an estimate of the form $N\leq Z+C$, where $C>0$ is independent of $Z>0$. For molecules, one seeks an analogous bound with a constant that depends only on the number of nuclei and is independent of their charges and positions (cf. \cite{BSS}). One also seeks uniform bounds on the radius outside which a prescribed number of electrons remain. A related question (cf. \cite{SolovejHF,Solovej2016}) concerns a uniform bound on the energy required to remove a prescribed number of electrons from the system.

In many-body Schr\"odinger theory, this problem has been extensively studied (cf. \cite{FS1990,HPS2025,LiebIonization,LSST,Nam2012,SSS1990,Sigal1982}). For a fermionic atom, it is known (cf. \cite{FS1990,HPS2025,LiebIonization,Nam2012,SSS1990}) that 
\begin{equation*}
N\leq \min\{2Z+1,1.1185Z+4Z^{1/3},1.22Z+3Z^{1/3},Z+CZ^{5/7}+C\}
\end{equation*}
Very recently, OpenAI (cf. \cite{OpenAICoulomb}) considered the full nonrelativistic Coulomb Hamiltonian with $K$ fixed nuclei at distinct positions, each with charge $Z_j\geq1$. They proved that $E_n<E_{n-1}$ implies that $n\leq Z+CK$ for every integer $n\geq1$, where $Z=\sum_{j=1}^KZ_j$, $E_n$ denotes the ground-state energy of the molecule with $n$ electrons, and $C>0$ is a universal constant.

The ionization conjecture has also been studied in models that approximate the many-electron Schr\"odinger theory of atoms and molecules. The Thomas--Fermi (TF) model introduced by Thomas and Fermi (cf. \cite{Fermi,Thomas}) describes the electronic energy in terms of the electron density. In \cite{LiebSimon}, Lieb and Simon proved that the ground-state energy of atoms and molecules in many-electron Schr\"odinger theory is asymptotic to the corresponding TF energy in a suitable large-charge limit. Benguria, Brezis and Lieb developed (cf.  \cite{BBL}) the variational theory of the Thomas--Fermi--von Weizs\"acker (TFW) functional, and Benguria and Lieb established (cf. \cite{BenguriaLieb}) excess-charge bounds of the TFW atoms and molecules. Solovej subsequently proved in \cite{SolovejTFW} that the electron densities in the TFW model converge to a limiting density as the nuclear charges tend to infinity. In the atomic Hartree--Fock theory, Solovej proved (cf. \cite{SolovejHF}) a uniform excess-charge bound, together with uniform bounds on the ionization energy and electronic-radius estimates for neutral atoms.

In this paper, we consider the molecular Thomas--Fermi--Dirac--von Weizs\"acker (TFDW) model, which extends TF theory by adding Dirac's local exchange energy (cf. \cite{Dirac}) and von Weizs\"acker's gradient correction (cf. \cite{Weizsacker}). More precisely, we consider the minimization problem
\begin{equation}\label{eq:inf}
\begin{aligned}
I_V(N):=\inf_{\rho\in\mathcal A_N}\mathcal E_V(\rho),\ \ \mathcal A_N:=\Big\{\rho\geq0:\sqrt\rho\in H^1(\R^3),\ \int_{\R^3}\rho dx=N\Big\}
\end{aligned}
\end{equation}
where the energy functional $\mathcal E_V$ is defined by
\begin{equation}\label{eq:functional}
\begin{aligned}
\mathcal E_V(\rho):=c^{\mathrm W}\int_{\R^3}|\nabla\sqrt\rho|^2 dx+c^{\TF}\int_{\R^3}\rho^{5/3} dx-c^{\mathrm D}\int_{\R^3}\rho^{4/3} dx+D(\rho)-\int_{\R^3}V\rho dx.
\end{aligned}
\end{equation}
Here $D(f):=\frac12\iint_{\R^3\times\R^3} \frac{f(x)f(y)}{|x-y|} dx dy$, the fixed positive constants $c^{\mathrm W},c^{\TF},c^{\mathrm D}$ are chosen appropriately (cf. \cite{LiebReview,Schwinger}), and the nuclear potential is given by
\begin{equation}\label{eq:V}
V(x):=\sum_{j=1}^K\frac{Z_j}{|x-R_j|},\ \  Z_j>0,\ \  R_j\in\R^3,\ \  Z:=\sum_{j=1}^K Z_j.
\end{equation}
The masses $N>0$ and charges $Z>0$ are not necessarily integers. Coincident centers in \eqref{eq:V} are understood as a single center with the sum of their charges. The variational problem defined by the TFDW functional \eqref{eq:functional} differs from that for the full many-electron Coulomb Hamiltonian studied in \cite{OpenAICoulomb}. Therefore, the results of \cite{OpenAICoulomb} do not directly apply to the minimization problem \eqref{eq:inf}.

Lions proved in 1987 that the infimum $I_V(N)$ is attained (cf. \cite{Lions}) for $0<N\leq Z$. In \cite{LeBris}, Le Bris extended this existence result to $Z<N<Z+\varepsilon$ for some $\varepsilon>0$. Lu and Otto proved (cf. \cite{LuOtto}) the nonexistence of minimizers for \eqref{eq:inf} with $V\equiv0$ at sufficiently large mass. In \cite{NVdB}, Nam and Van Den Bosch showed that the minimization problem \eqref{eq:inf} has no minimizer for sufficiently large $N>0$ if the total nuclear charge $Z>0$ is sufficiently small. Aguirre Salazar, Alama and Bronsard (cf. \cite{AASAB}) established the finite-profile splitting of molecular TFDW minimizing sequences.

Regarding the ionization conjecture, Frank, Nam and Van Den Bosch proved in \cite{FNV} that every atomic TFDW minimizer of mass $N>0$ and nuclear charge $Z>0$ satisfies $N\leq Z+C$, where $C>0$ is independent of $Z>0$. Our first theorem gives a uniform bound on the excess charge $N-Z$ with a constant depending only on the number of nuclei.

\begin{theorem}\label{thm:1.1}
Let $I_V(N)$ be defined by \eqref{eq:inf}. For every $K\in\mathbb N^+$, there exists $C=C(K)>0$ such that for all positive nuclear charges $Z_i>0$, all nuclear positions $R_i\in\R^3$ and every $N>0$, if $I_V(N)$ has a minimizer, then we have
\begin{equation}\label{eq:unibound}
N\leq Z+C.
\end{equation}
\end{theorem}

Theorem~\ref{thm:1.1} answers the molecular large-charge question posed in \cite{FNV}. The bound \eqref{eq:unibound} requires neither comparability of the nuclear charges nor a positive lower bound on the distances between nuclei. These two assumptions are imposed in \cite[Theorem~1.1]{Goto} for molecular reduced Hartree--Fock theory. We expect that our method can be adapted to establish the bound $N\leq Z+C(K)$ of minimizers for the molecular Hartree--Fock functional.

Let $\rho$ be a minimizer of \eqref{eq:inf}, and let $\rho_{\TF}$ be the neutral molecular TF minimizer corresponding to the potential $V$ defined in \eqref{eq:V}. More precisely, $\rho_{\TF}$ minimizes $c^{\TF}\int_{\R^3}f^{5/3} dx+D(f)-\int_{\R^3}Vf dx$ among nonnegative $f\in L^1(\R^3)\cap L^{5/3}(\R^3)$ satisfying $\int_{\R^3}f dx=Z$. For $K=1$, the proof in \cite{FNV} estimates $|\int_{B(R_1,r)}(\rho-\rho_{\TF}) dx|$ and $\int_{\R^3\setminus B(R_1,r)}\rho dx$ at a suitable radius $r>0$. In a molecule, the balls $B(R_i,r)$ may overlap, and the ratio $\max_jZ_j/\min_jZ_j$ can be arbitrarily large. The atomic TF lower bound used in the comparison requires $Z_i r^3\geq1$. We therefore select at each radius $r$ the nuclei satisfying this condition. We adapt the atomic argument in the following three steps.

In the first step, we partition the region outside the selected balls according to the nearest nucleus and adapt the cutting and averaging argument in \cite[Lemma~3.2]{FNV}. Each cut keeps all the electrons inside the union of the balls on the same side. The energy comparison therefore includes the Coulomb repulsion between these electrons and those on the other side, even when the balls overlap. This gives an exterior mass estimate for every $r>0$, uniformly in the nuclear positions; see Lemmas~\ref{lem:locale} and \ref{lem:exterior2} for more details.

In the second step, we consider an auxiliary TF problem that includes the Coulomb potentials of the nuclei not yet selected. In Section~\ref{eq:section}, we adapt the arguments in \cite[Lemmas~6.8 and 6.9]{FNV} to compare the auxiliary TF potential with the neutral molecular TF potential and to obtain a localized energy estimate. In Proposition~\ref{iter:enlarged-screening}, we combine these estimates to compare the TFDW and neutral TF screened potentials for every set of nuclei containing the selected ones.

In the final step, we extend the screened-potential comparison to larger radii and add further nuclei as soon as their charges satisfy $Z_i r^3\geq1$. The estimates remain uniform in the charge ratios. As shown in Lemma~\ref{iter:finite-screening}, after finitely many steps we obtain the comparison for the full set of nuclei for every $0<r\leq D_*$, where $D_*>0$ depends only on $K$. At $r=D_*$, this comparison bounds the difference between the masses of $\rho$ and $\rho_{\TF}$ inside the union of the balls. Combining this bound with the estimate for the mass of $\rho$ outside the union proves \eqref{eq:unibound}.

For the following comparisons, assume that $N\geq Z$. For $r\geq0$, we define
\begin{equation}\label{eq:domain1}
\begin{aligned}
d(x):=&\min_{1\leq j\leq K}|x-R_j|,\ \  \Omega_r:=\{x\in\R^3:d(x)\leq r\},\ \  A_r:=\R^3\setminus\Omega_r.
\end{aligned}
\end{equation}
We define the corresponding screened potentials by
\begin{equation}\label{eq:screen1}
\Phi_r:=V-(\ind_{\Omega_r}\rho)*\coul,\ \  \Phi_r^{\TF}:=V-(\ind_{\Omega_r}\rho_{\TF})*\coul.
\end{equation}

\begin{theorem}\label{thm:screen}
For every $K\in\mathbb N^+$, there exists $C=C(K)>0$ such that for every $r>0$, we have
\begin{equation}\label{eq:screebound}
\sup_{\partial\Omega_r}|\Phi_r-\Phi_r^{\TF}| \leq C(1+r^{-4+10^{-5}}),
\end{equation}
and
\begin{equation}\label{eq:dmass}
\Big|\int_{A_r}(\rho-\rho_{\TF}) dx\Big| \leq C(1+r^{-3+10^{-5}}).
\end{equation}
\end{theorem}

The estimate \eqref{eq:screebound} extends the atomic screened-potential comparison in \cite[Theorem~1.2]{FNV} to molecules, uniformly in the nuclear charges and positions. For $0<r\leq D_*$, Lemma~\ref{iter:finite-screening} gives \eqref{eq:screebound} for the full set of nuclei. From \eqref{eq:screenm}, we then obtain a bound on $\big|\int_{\Omega_r}(\rho-\rho_{\TF}) dx\big|$. Together with $0\leq N-Z\leq C(K)$ from Theorem~\ref{thm:1.1}, this yields \eqref{eq:dmass}. For $r>D_*$, the exterior mass bounds give \eqref{eq:dmass}. To prove \eqref{eq:screebound} for $r>D_*$, we combine the screened-potential bound at $D_*/2$ with the exterior mass and $L^{5/3}$ estimates.

For $0\leq f\in L^1(\R^3)$ and $0<\kappa<\int_{\R^3}f dx$, we define
\begin{equation}\label{eq:Rf}
R_f(\kappa) :=\sup\Big\{r\geq0:\int_{A_r}f dx=\kappa\Big\},
\end{equation}
where $A_r$ is as in \eqref{eq:domain1}. Denote also $R_{\TF}:=R_{\rho_{\TF}}$.

\begin{theorem}\label{thm:radius}
For every $K\in\mathbb N^+$, there exist constants $A_K>1$ and $C_K,c_K,\kappa_0>0$, depending only on $K$, such that for $\kappa\geq\kappa_0$ and $Z\geq A_K\kappa$, we have
\begin{equation}\label{eq:radiusb}
c_K\kappa^{-1/3}\leq R_\rho(\kappa),R_{\TF}(\kappa) \leq C_K\kappa^{-1/3},
\end{equation}
and
\begin{equation}\label{eq:radiusd}
\left|R_\rho(\kappa)-R_{\TF}(\kappa)\right| \leq C_K\kappa^{-1/3-10^{-5}/3}.
\end{equation}
\end{theorem}

The atomic radius estimate in \cite[Theorem~1.3]{FNV} uses an explicit reference radius. For molecules, we compare $R_\rho(\kappa)$ with the corresponding neutral TF radius $R_{\TF}(\kappa)$. For fixed $K$, we deduce from \eqref{eq:radiusb} and \eqref{eq:radiusd} that $R_\rho(\kappa)/R_{\TF}(\kappa)\to1$ as $\kappa\to\infty$, uniformly in the nuclear charges and positions satisfying $Z\geq A_K\kappa$.

To derive \eqref{eq:radiusd} from the mass comparison \eqref{eq:dmass}, we establish that
\begin{equation*}
\int_{A_s\setminus A_t}\rho_{\TF} dx\geq c_K(t-s)t^{-4}\text{ \ for \ }0<s<t\text{ \ with \ }s^3\max_jZ_j\geq1.
\end{equation*}
We prove this bound by integrating the neutral atomic TF density estimate along rays from a nucleus of maximal charge. This argument also applies when the nuclear balls overlap.

By an energy comparison, we establish the following uniform upper bound on the energy required to remove electronic mass from a minimizer of \eqref{eq:inf}.

\begin{theorem}\label{thmm:ener}
There is a constant $C=C(K)>0$ such that if $I_V(N)$ has a minimizer with $N\geq Z$, then we have
\begin{equation}\label{eq:ener1}
0\leq I_V(N-\kappa)-I_V(N) \leq C(1+\kappa^{7/3})\text{ \ for \  }0<\kappa\leq N.
\end{equation}
\end{theorem}

The difference in \eqref{eq:ener1} is the energy required to remove $\kappa$ units of electronic mass with the nuclei fixed. Solovej proved a uniform upper bound on the first ionization energy of neutral Hartree--Fock atoms in \cite[Theorem~3.8]{SolovejHF}. For molecular TFDW minimizers, we use Theorem~\ref{thm:radius} to choose a cutoff that removes at least $\kappa$ units of mass and Theorem~\ref{thm:screen} to control the nuclear attraction in the energy comparison. The argument does not require $I_V(N-\kappa)$ to have a minimizer.

The paper is organized as follows. In Section~\ref{sec:prelim}, we establish some preliminary estimates. Section~\ref{eq:section} is devoted to studying an exterior TF problem. In Section~\ref{iter:section}, we finally prove Theorems~\ref{thm:1.1}--\ref{thmm:ener}.

\section{Localization and Initial Thomas--Fermi Comparison}\label{sec:prelim}

In this section, we establish some preliminary estimates. Unless otherwise stated, $C>0$ denotes a positive constant depending only on $K$ and may change from line to line.

Set $d_j(x):=|x-R_j|$. For a nonempty set $I\subset\{1,\ldots,K\}$ and $r>0$, we define
\begin{equation}\label{eq:actived}
d_I(x):=\min_{i\in I}d_i(x),\ \ A_r^I:=\{x\in\R^3:d_I(x)>r\}.
\end{equation}
Sums over $j\notin I$ are understood to be zero when $I=\{1,\ldots,K\}$.
We also define
\begin{equation}\label{eq:active-inner}
\Omega_r^I:=\R^3\setminus A_r^I=\bigcup_{i\in I}\overline{B(R_i,r)}.
\end{equation} 
For a TFDW minimizer $\rho$, we define the screened potentials associated with $\Omega_r^I$ by
\begin{equation}\label{eq:screen}
\Phi_r^I:=V-(\ind_{\Omega_r^I}\rho)*\coul,\ \  \Phi_r^{I,\TF}:=V-(\ind_{\Omega_r^I}\rho_{\TF})*\coul,
\end{equation}
so that $\Phi_r^I-\Phi_r^{I,\TF}\equiv-[\ind_{\Omega_r^I}(\rho-\rho_{\TF})]*\coul$ is continuous in $\R^3$ and satisfies $\Delta(\Phi_r^I-\Phi_r^{I,\TF})=0$ in $A_r^I$. Recall from \cite[Lemma~5]{NVdB} that for every $f\in\mathcal A_M$ with $0\leq M\leq N$, where $\mathcal A_M$ is as in \eqref{eq:inf}, we have
\begin{equation}\label{eq:massm}
I_V(N)\leq\mathcal E_V(f),
\end{equation}
and we also have
\begin{equation}\label{eq:energym}
I_V(N)\leq I_V(M)\leq0\text{ \ for \  }0\leq M\leq N, \ \  I_V(0)=0.
\end{equation}

We adapt the averaged-cut argument in \cite[Lemma~3.2]{FNV}, based on \cite{FKN}, to estimate the mass of $\rho$ in a subset of $A_r^I$.

\begin{lemma}\label{lem:locale}
Let $I$ be nonempty and fix $r>0$ and $q_0\in\R^3$. Suppose that the Lipschitz function $\Theta:\R^3\to\R^3$ satisfies $\Theta\equiv0$ on $\Omega_r^I$ and $\Theta(x)\equiv x-q_0$ on $F\subset A_r^I$. Set $S:=\{x\in\R^3:\Theta(x)\ne0\}$ and assume that for some constants $L,Q\geq0$,
\begin{equation}\label{eq:lip}
\Lip(\Theta)\leq L,\ \  |\Theta|[\Phi_r^I]_+\leq Q\text{ \ on \  }S.
\end{equation}
Here and below, we use the notation $[\cdot]_+:=\max\{\cdot,0\}$. Then for every $t>0$, we have
\begin{equation}\label{eq:loce}
\frac18\left(\int_F\rho dx\right)^2 \leq C(t+L^2/t)\int_S\rho dx+\frac Q4\int_S\rho dx +2tD(\ind_F\rho).
\end{equation}
\end{lemma}

\begin{proof}
Let $\chi,\eta\in W^{1,\infty}(\R^3;[0,1])$ satisfy that $\chi^2+\eta^2\equiv1$ in $\R^3$ and $\chi\equiv1$ on $\Omega_r^I$. Set $f:=\eta^2\rho$. Since $\chi^2\rho\equiv\rho$ on $\Omega_r^I$, we deduce from \eqref{eq:screen} that
\begin{equation*}
\int_{\R^3}\big(V-(\chi^2\rho)*\coul\big)f dx
=\int_{A_r^I}\Phi_r^I f dx-\int_{A_r^I}f\big[(\ind_{A_r^I}\chi^2\rho)*\coul\big] dx.
\end{equation*}
By the IMS formula (cf. \cite[Theorem~3.2]{CFKS87}) and the estimates of the TF and Dirac terms in \cite[proof of Lemma~3.1]{FNV}, we obtain that
\begin{equation}\label{eq:binding}
\begin{aligned}
&\mathcal E_V(\chi^2\rho)+\mathcal E_0(f)-\mathcal E_V(\rho)+\int_{A_r^I}f\big[(\ind_{A_r^I}\chi^2\rho)*\coul\big] dx\\
\leq&c^{\mathrm W}\int_{\R^3}(|\nabla\chi|^2+|\nabla\eta|^2)\rho dx+C\int_{\{0<\chi<1\}}\rho dx+\int_{A_r^I}\Phi_r^I f dx.
\end{aligned}
\end{equation}

Let $g_1,g_2\in C^\infty(\R;[0,1])$ satisfy that $g_1^2+g_2^2\equiv1$ in $\R$, $g_1\equiv1$ in $(-\infty,0]$ and $g_1\equiv0$ in $[1,\infty)$. Define for $\omega\in\Sph:=\{\omega\in\R^3:|\omega|=1\}$,
\begin{equation}\label{eq:planecut}
\begin{aligned}
\chi_i(x):=g_i((\omega\cdot\Theta(x)-\ell)/t),\ \ T_{\omega,\ell}:=\{x\in\R^3:\ell<\omega\cdot\Theta(x)<\ell+t\},
\end{aligned}
\end{equation}
where $i=1,2$ and $\ell>0$, so that $|\nabla\chi_i|\leq C L t^{-1}\ind_{T_{\omega,\ell}}$ almost everywhere. By the binding inequality in \cite[Lemma~5]{NVdB}, we have $\mathcal E_V(\chi_1^2\rho)+\mathcal E_0(\chi_2^2\rho)\geq\mathcal E_V(\rho)$. For $x,y\in F$ satisfying $\omega\cdot(y-q_0)\leq\ell\leq\omega\cdot(x-q_0)-t$, we have $\chi_2(x)=\chi_1(y)=1$ by \eqref{eq:planecut}. Applying \eqref{eq:binding} with $\chi=\chi_1$ and $\eta=\chi_2$, we thus obtain that
\begin{equation}\label{eq:2.22}
\begin{aligned}
&\iint_{\substack{x,y\in F\\\omega\cdot(y-q_0)\leq\ell\leq\omega\cdot(x-q_0)-t}}\frac{\rho(x)\rho(y)}{|x-y|} dx dy\\
\leq&\iint_{\R^3\times A_r^I}\frac{\chi_2(x)^2\rho(x)\chi_1(y)^2\rho(y)}{|x-y|} dx dy\leq C(1+L^2t^{-2})\int_{T_{\omega,\ell}}\rho dx+\int_{\R^3}\chi_2^2\rho\Phi_r^I dx.
\end{aligned}
\end{equation}

It follows from \eqref{eq:planecut} that
\begin{equation}\label{eq2.23}
\begin{aligned}
&\int_0^\infty\ind_{T_{\omega,\ell}}(x)d\ell =[\omega\cdot\Theta(x)]_+-[\omega\cdot\Theta(x)-t]_+\leq t\ind_S(x),\\
&\int_{\Sph}\int_0^\infty\ind_{\{\ell<\omega\cdot\Theta(x)\}}d\ell\frac{d\omega}{4\pi} =\int_{\Sph}[\omega\cdot\Theta(x)]_+\frac{d\omega}{4\pi} =\frac{|\Theta(x)|}{4}.
\end{aligned}
\end{equation}
Since $0\leq\chi_2(x)^2\leq\ind_{\{\ell<\omega\cdot\Theta(x)\}}$, we deduce from \eqref{eq:lip} and \eqref{eq2.23} that
\begin{equation}\label{eq:2.24}
\begin{aligned}
&\int_{\Sph}\int_0^\infty\Big(C(1+L^2t^{-2})\int_{T_{\omega,\ell}}\rho dx+\int_{\R^3}\chi_2^2\rho\Phi_r^I dx\Big)d\ell\frac{d\omega}{4\pi}\\
\leq&C(t+L^2/t)\int_S\rho dx+\frac14\int_S|\Theta|\rho[\Phi_r^I]_+ dx
\leq C(t+L^2/t)\int_S\rho dx+\frac Q4\int_S\rho dx.
\end{aligned}
\end{equation}
For $A,B\in\R$ and $t\geq0$, we define
\begin{equation*}
\Lambda_t(A,B):=\int_0^\infty\ind_{\{B\leq\ell\leq A-t\}}d\ell=[A-t-\max\{B,0\}]_+.
\end{equation*}
Note that $\Lambda_t(A,B)\geq\Lambda_0(A,B)-t$ and $\Lambda_0(A,B)+\Lambda_0(-B,-A)=[A-B]_+$. We then have
\begin{equation}\label{eq:2.25}
\begin{aligned}
\Lambda_t(A,B)+\Lambda_t(-B,-A) \geq\Lambda_0(A,B)+\Lambda_0(-B,-A)-2t=[A-B]_+-2t.
\end{aligned}
\end{equation}
Since $\int_{\Sph}[\omega\cdot(x-y)]_+\frac{d\omega}{4\pi}=\frac{|x-y|}{4}$, we obtain from \eqref{eq:2.25} and the symmetry $(\omega,x,y)\mapsto(-\omega,y,x)$ that
\begin{equation}\label{eq:2.26}
\begin{aligned}
&\int_{\Sph}\int_0^\infty\iint_{\substack{x,y\in F\\B\leq\ell\leq A-t}}\frac{\rho(x)\rho(y)}{|x-y|} dx dy d\ell\frac{d\omega}{4\pi}\\
=&\int_{\Sph}\iint_{F\times F}\Lambda_t(A,B)\frac{\rho(x)\rho(y)}{|x-y|} dx dy\frac{d\omega}{4\pi}\\
=&\frac12\int_{\Sph}\iint_{F\times F}\bigl(\Lambda_t(A,B)+\Lambda_t(-B,-A)\bigr)\frac{\rho(x)\rho(y)}{|x-y|} dx dy\frac{d\omega}{4\pi}\\
\geq&\int_{\Sph}\iint_{F\times F}\left(\frac12[\omega\cdot(x-y)]_+-t\right)\frac{\rho(x)\rho(y)}{|x-y|} dx dy\frac{d\omega}{4\pi}\\
=&\iint_{F\times F}\left(\frac{|x-y|}{8}-t\right)\frac{\rho(x)\rho(y)}{|x-y|} dx dy=\frac18\left(\int_F\rho dx\right)^2-2tD(\ind_F\rho),
\end{aligned}
\end{equation}
where $A:=\omega\cdot(x-q_0)$ and $B:=\omega\cdot(y-q_0)$. Integrating \eqref{eq:2.22} over $0<\ell<\infty$ and $\omega\in\Sph$ with respect to $d\omega/(4\pi)$, we conclude from \eqref{eq:2.24} and \eqref{eq:2.26} that \eqref{eq:loce} holds. This completes the proof of Lemma~\ref{lem:locale}.
\end{proof}

Define the harmonic function $h_I$ in $A_r^I$ by
\begin{equation}\label{eq:harmoniccom}
\begin{aligned}
h_I(x):=|I|^{-1}\sum_{i\in I}|x-R_i|^{-1}.
\end{aligned}
\end{equation}
Applying Lemma~\ref{lem:locale}, we establish a bound on the exterior mass $\int_{A_r^I}\rho dx$.

\begin{lemma}\label{lem:exterior2}
Let $I$ be nonempty and fix $r>0$. Let $h_I$ and $\Phi_r^I$ be defined in \eqref{eq:harmoniccom} and \eqref{eq:screen}, respectively. Suppose that for some constant $Q\geq0$,
\begin{equation}\label{eq:2.27}
\Phi_r^I(x)\leq Qh_I(x)+\sum_{j\notin I}\frac{Z_j}{|x-R_j|}\text{ \ for \  }x\in A_r^I.
\end{equation}
Then we have for every $t>0$,
\begin{equation}\label{eq:2.28}
\begin{aligned}
\int_{A_r^I}\rho dx \leq&C\int_{A_r^I\setminus A_{3r/2}^I}\rho dx+C\big(Q+\sum_{i\notin I}Z_i+t+4t^{-1}\big)\\
&\ \ \ +Ct^{6/5}\Big(\int_{A_{3r/2}^I}\rho^{5/3} dx\Big)^{3/5}.
\end{aligned}
\end{equation}
\end{lemma}

\begin{proof}
If $\int_{A_r^I}\rho dx=0$, the assertion is immediate. We may therefore assume that $\int_{A_r^I}\rho dx>0$. Let $d(x)=\min_{1\leq i\leq K}d_i(x)$ be as in \eqref{eq:domain1}. Define for $1\leq j\leq K$,
\begin{equation}\label{eq:map2}
w(x):=\min\{d(x),3[d_I(x)-r]_+\},\ \ 
\Theta_j(x):=w(x)\frac{x-R_j}{d_j(x)},
\end{equation}
where $ \Theta_j(R_j):=0$. We then have $\Lip(w)\leq3$. Since $0\leq w\leq d_j$, direct calculations give that for $x,y\neq R_j$
\begin{equation*}
\begin{aligned}
|\Theta_j(x)-\Theta_j(y)|^2
&=|w(x)-w(y)|^2+\frac{w(x)w(y)}{d_j(x)d_j(y)}
\bigl(|x-y|^2-|d_j(x)-d_j(y)|^2\bigr)\\
&\leq |w(x)-w(y)|^2+|x-y|^2.
\end{aligned}
\end{equation*}
Since $|\Theta_j(x)|=w(x)\leq|x-R_j|$ and $\Theta_j(R_j)=0$, the same bound also holds when $x=R_j$ or $y=R_j$. We thus conclude that
\begin{equation}\label{eq:2.32}
|\Theta_j(x)-\Theta_j(y)|\leq |w(x)-w(y)|+|x-y|\leq4|x-y|\text{ \ for \ }x,y\in\R^3.
\end{equation}

Define for $j\in\{1,\cdots,K\}$
\begin{equation}\label{eq:2.35}
\Gamma_j:=\big\{x\in\R^3:d_j(x)=\min_{1\leq i\leq K}d_i(x)\big\}\setminus\bigcup_{i<j}\{x\in\R^3:d_i(x)=d_j(x)\}.
\end{equation}
Set $F_j:=\Gamma_j\cap A_{3r/2}^I$. Then the sets $F_j$ partition $A_{3r/2}^I$. On $F_j$, we have $d_j\leq d_i$ for every $i\in\{1,\cdots,K\}$ and $d_I>3r/2$. Hence $3(d_I-r)\geq d_I\geq d=d_j$ on $F_j$, and it yields from \eqref{eq:map2}  that
\begin{equation}\label{eq:2.36}
\Theta_j(x)\equiv x-R_j\text{ \ on \  }F_j.
\end{equation}
Since $|\Theta_j|=w\leq d_i$ for every $i$, it follows from \eqref{eq:harmoniccom} that $wh_I=|I|^{-1}\sum_{i\in I}w/d_i\leq1$ on $\{w>0\}$. We therefore deduce from \eqref{eq:2.27} that
\begin{equation}\label{eq:2.37}
|\Theta_j|[\Phi_r^I]_+\leq Qwh_I+\sum_{i\notin I}Z_i w d_i^{-1}\leq Q+\sum_{i\notin I}Z_i\text{ \ on \ }\{w>0\}.
\end{equation}

Applying Lemma~\ref{lem:locale} with $F=F_j$, $q_0=R_j$ and $L=4$, we obtain from \eqref{eq:loce}, \eqref{eq:2.32}, \eqref{eq:2.36} and \eqref{eq:2.37} that for $t>0$,
\begin{equation}\label{eq:2.38}
\begin{aligned}
\Big(\int_{F_j}\rho dx\Big)^2\leq C\big(Q+\sum_{i\notin I}Z_i+t+4t^{-1}\big)\int_{\{x\in\R^3:\Theta_j(x)\ne0\}}\rho dx +CtD(\ind_{F_j}\rho).
\end{aligned}
\end{equation}
Since $\{x\in\R^3:\Theta_j(x)\ne0\}\subset A_r^I$ and the sets $F_j$ are disjoint, we have
\begin{equation}\label{eq:2.39}
\begin{aligned}
\sum_j\int_{\{x\in\R^3:\Theta_j(x)\ne0\}}\rho dx\leq K\int_{A_r^I}\rho dx,\ \ \sum_jD(\ind_{F_j}\rho)\leq D(\ind_{A_{3r/2}^I}\rho).
\end{aligned}
\end{equation}
We obtain by the Hardy--Littlewood--Sobolev inequality \cite[Theorem~4.3]{LiebLoss} that  for all $f\in L^1(\R^3)\cap L^{5/3}(\R^3)$,
\begin{equation}\label{eq:HLSe}
\begin{aligned}
D(f)\leq&C\|f\|_{6/5}^2\leq C\|f\|_1^{7/6}\|f\|_{5/3}^{5/6} =C\Big(\int_{\R^3}|f| dx\Big)^{7/6}\Big(\int_{\R^3}|f|^{5/3} dx\Big)^{1/2},
\end{aligned}
\end{equation}
where $D(\cdot)$ is defined in \eqref{eq:functional}. 

By the Cauchy--Schwarz inequality, we obtain from \eqref{eq:2.38}, \eqref{eq:2.39} and \eqref{eq:HLSe} that

\begin{align}
\Big(\int_{A_{3r/2}^I}\rho dx\Big)^2 =&\Big(\sum_j\int_{F_j}\rho dx\Big)^2\leq K\sum_j\Big(\int_{F_j}\rho dx\Big)^2\notag\\
\leq&C(Q+\sum_{i\notin I}Z_i+t+4t^{-1})\int_{A_r^I}\rho dx+CtD(\ind_{A_{3r/2}^I}\rho)\label{eq:2.40}\\
\leq&C(Q+\sum_{i\notin I}Z_i+t+4t^{-1})\int_{A_r^I}\rho dx+Ct\Big(\int_{A_{3r/2}^I}\rho dx\Big)^{7/6}\Big(\int_{A_{3r/2}^I}\rho^{5/3} dx\Big)^{1/2}\notag.
\end{align}
Since $A_r^I=A_{3r/2}^I\cup(A_r^I\setminus A_{3r/2}^I)$, we have
\begin{equation}\label{eq:2.41}
\begin{aligned}
\Big(\int_{A_r^I}\rho dx\Big)^2\leq&2\Big(\int_{A_r^I\setminus A_{3r/2}^I}\rho dx\Big)^2+2\Big(\int_{A_{3r/2}^I}\rho dx\Big)^2\\
\leq&2\Big(\int_{A_r^I\setminus A_{3r/2}^I}\rho dx\Big)\Big(\int_{A_r^I}\rho dx\Big)+2\Big(\int_{A_{3r/2}^I}\rho dx\Big)^2.
\end{aligned}
\end{equation}
By Young's inequality, we obtain from \eqref{eq:2.40} and \eqref{eq:2.41} that
\begin{equation*}
\begin{aligned}
&\int_{A_r^I}\rho dx\\
\leq&2\int_{A_r^I\setminus A_{3r/2}^I}\rho dx+C(Q+\sum_{i\notin I}Z_i+t+4t^{-1})+Ct\Big(\int_{A_r^I}\rho dx\Big)^{1/6}\Big(\int_{A_{3r/2}^I}\rho^{5/3} dx\Big)^{1/2}\\
\leq&2\int_{A_r^I\setminus A_{3r/2}^I}\rho dx+C(Q+\sum_{i\notin I}Z_i+t+4t^{-1})+\frac12\int_{A_r^I}\rho dx+Ct^{6/5}\Big(\int_{A_{3r/2}^I}\rho^{5/3} dx\Big)^{3/5},
\end{aligned}
\end{equation*}
where we have used $\int_{A_{3r/2}^I}\rho dx\leq\int_{A_r^I}\rho dx$. This thus completes the proof of Lemma~\ref{lem:exterior2}.
\end{proof}

We next prove a rough bound of $N$.

\begin{lemma}\label{lem:coarse}
Every TFDW minimizer $\rho$ of $I_V(N)$ defined in \eqref{eq:inf} satisfies that
\begin{equation}\label{eq:coarse-bounds}
\begin{aligned}
N\leq C(Z+1),\ \ \int_{\R^3}|\nabla\sqrt\rho|^2 dx+\int_{\R^3}\rho^{5/3} dx+D(\rho)\leq C(Z^{7/3}+Z+1).
\end{aligned}
\end{equation}
\end{lemma}

\begin{proof}
Let $0\leq f\in L^1(\R^3)\cap L^{5/3}(\R^3)$, so that $f\in L^{6/5}(\R^3)$. By Young's inequality, we obtain from  \cite[(6.3)]{FNV} that
\begin{equation}\label{eq:2.43}
\begin{aligned}
\int_{\R^3}Vf dx=\sum_{j=1}^K Z_j(f*\coul)(R_j)
&\leq CZ\Big(\int_{\R^3}f^{5/3} dx\Big)^{3/7}D(f)^{1/7}\\
&\leq\frac{c^{\TF}}2\int_{\R^3}f^{5/3} dx+\frac12D(f)+CZ^{7/3}.
\end{aligned}
\end{equation}
Since $c^{\mathrm D}u^{4/3}\leq(c^{\TF}/4)u^{5/3}+Cu$ for $u\geq0$, we derive from \eqref{eq:2.43} that  for $f\geq0$ satisfying $\sqrt f\in H^1(\R^3)$,
\begin{equation}\label{eq:2.44}
\begin{aligned}
\mathcal E_V(f)=&c^{\mathrm W}\int_{\R^3}|\nabla\sqrt f|^2 dx+\frac{c^{\TF}}2\int_{\R^3}f^{5/3} dx+\frac12D(f)-c^{\mathrm D}\int_{\R^3}f^{4/3} dx\\
&+\left(\frac{c^{\TF}}2\int_{\R^3}f^{5/3} dx+\frac12D(f)-\int_{\R^3}Vf dx\right)\\
\geq&c^{\mathrm W}\int_{\R^3}|\nabla\sqrt f|^2 dx+\frac{c^{\TF}}4\int_{\R^3}f^{5/3} dx+\frac12D(f)-CZ^{7/3}-C\int_{\R^3}f dx.
\end{aligned}
\end{equation}
Together with $I_V(N)\leq0$ in \eqref{eq:energym}, this implies that
\begin{equation}\label{eq:2.45}
\int_{\R^3}|\nabla\sqrt\rho|^2 dx+\int_{\R^3}\rho^{5/3} dx+D(\rho) \leq C(Z^{7/3}+N).
\end{equation}

Since $\rho\in L^1(\R^3)\cap L^{5/3}(\R^3)$, we have
\begin{equation}\label{eq:limit}
\int_{A_r^I}\rho dx\to N,\ \  \int_{A_r^I\setminus A_{3r/2}^I}\rho dx\to0\text{ \ and \  }\int_{A_{3r/2}^I}\rho^{5/3} dx\to\int_{\R^3}\rho^{5/3} dx\text{ \ as \  }r\to0^+.
\end{equation}
For $I=\{1,\ldots,K\}$, we obtain from \eqref{eq:V}, \eqref{eq:screen} and \eqref{eq:harmoniccom} that $\Phi_r^I\leq V\leq KZh_I$. Applying Lemma~\ref{lem:exterior2} with $Q=KZ$ and letting $r\to0^+$, we conclude from \eqref{eq:2.28}, \eqref{eq:2.45} and \eqref{eq:limit} that for $t>0$,
\begin{equation}\label{eq:2.47}
N\leq C(Z+t+t^{-1})+Ct^{6/5}(Z^{7/3}+N)^{3/5}.
\end{equation}
Taking $t=(Z+1)^{-1/3}>0$ in \eqref{eq:2.47}, we then deduce by Young's inequality that
\begin{equation*}
N\leq C(Z+1)+C(Z+1)^{-2/5}N^{3/5} \leq C(Z+1)+\frac12N+C(Z+1)^{-1}.
\end{equation*}
This proves the mass estimate in \eqref{eq:coarse-bounds}, and substituting it into \eqref{eq:2.45} yields the energy estimate in \eqref{eq:coarse-bounds}. This completes the proof of Lemma~\ref{lem:coarse}.
\end{proof}

We next give a quantitative form of the convexity of $z\mapsto z^{5/3}$ on $[0,\infty)$.

\begin{lemma}\label{lem:tf-convexity}
We have for all nonnegative $u,v\in L^{5/3}(\R^3)$,
\begin{equation}\label{eq:tf-convexity}
\|u-v\|_{5/3}^2\leq\frac95\|u+v\|_{5/3}^{1/3}\int_{\R^3}\Big[u^{5/3}-v^{5/3}-\frac53v^{2/3}(u-v)\Big] dx.
\end{equation}
\end{lemma}

\begin{proof}
By Taylor's formula, we have for $p,q\geq0$ satisfying $p+q>0$,
\begin{equation}\label{eq:325}
\begin{aligned}
p^{5/3}-q^{5/3}-\frac53q^{2/3}(p-q)=&\frac{10}{9}(p-q)^2\int_0^1\frac{1-t}{((1-t)q+tp)^{1/3}} dt\\
\geq&\frac59\frac{(p-q)^2}{(p+q)^{1/3}}.
\end{aligned}
\end{equation}
H\"older's inequality then gives that
\begin{equation*}
\int_{\R^3}|u-v|^{5/3} dx\leq\left(\int_{\R^3}\frac{|u-v|^2}{(u+v)^{1/3}} dx\right)^{5/6}\left(\int_{\R^3}(u+v)^{5/3} dx\right)^{1/6},
\end{equation*}
where we take the quotient to be zero when $u=v=0$. Together with \eqref{eq:325}, this completes the proof of Lemma~\ref{lem:tf-convexity}.
\end{proof}

We recall some results of TF theory. For the potential $V$ defined in \eqref{eq:V}, consider the TF functional
\begin{equation}\label{eq:tff}
\mathcal F_V(f):=c^{\TF}\int_{\R^3}f^{5/3} dx+D(f)-\int_{\R^3}Vf dx.
\end{equation}
The minimization problem $\inf\left\{\mathcal F_V(f):0\leq f\in L^1(\R^3)\cap L^{5/3}(\R^3),\ \int_{\R^3}f dx=Z\right\}$ admits a unique neutral minimizer $\rho_{\TF}$ with zero chemical potential (cf. \cite[Theorems~II.10 and II.20]{LiebSimon}). By \cite[Theorem~IV.3]{LiebSimon}, with $A_{\TF}:=(3/(5c^{\TF}))^{3/2}$, we have
\begin{equation}\label{eq:tfe}
\begin{aligned}
\rho_{\TF}= A_{\TF}\phi_{\TF}^{3/2},\ \ \phi_{\TF}:=V-\rho_{\TF}*\coul>0\text{ \ in \  }\R^3\setminus\{R_1,\ldots,R_K\}.
\end{aligned}
\end{equation}
It follows from \eqref{eq:tfe} that
\begin{equation}\label{eq:tfpde}
-\Delta\phi_{\TF}+k_{\TF}\phi_{\TF}^{3/2}=4\pi\sum_jZ_j\delta_{R_j}\text{ \ in \  }\R^3,\ \ k_{\TF}:=4\pi A_{\TF}.
\end{equation}
By \cite[Theorem~IV.5]{LiebSimon}, we have $\phi_{\TF},\rho_{\TF}\in C^\infty(\R^3\setminus\{R_1,\ldots,R_K\})$. Let $\phi_j^{\rm at}$ and $\rho_j^{\rm at}$ be the neutral atomic TF potential and density corresponding to $V=Z_j/|x-R_j|$. By \cite[Theorems~IV.8, V.5 and V.12]{LiebSimon}, we have
\begin{equation}\label{eq:acomparison}
\max_j\phi_j^{\rm at}\leq\phi_{\TF}\leq\sum_j\phi_j^{\rm at},\ \  \phi_j^{\rm at}\leq C\min\{Z_jd_j^{-1},d_j^{-4}\}.
\end{equation}
By atomic scaling and the atomic TF estimates in \cite[Theorem~4.2]{FNV}, we have
\begin{equation}\label{eq:aestimates}
\begin{aligned}
\phi_j^{\rm at}(x)\geq c d_j(x)^{-4}\text{ \ for \  }Z_jd_j(x)^3\geq1,\ \ \int_{\R^3}\rho_j^{\rm at} dx=Z_j,\ \ \int_{\R^3}(\rho_j^{\rm at})^{5/3} dx \leq CZ_j^{7/3},
\end{aligned}
\end{equation}
where $d_j(x)=|x-R_j|$. By H\"older's inequality, we obtain from \eqref{eq:tfe} and \eqref{eq:acomparison} that
\begin{equation}\label{eq:tfdcom}
\begin{aligned}
\rho_{\TF}\leq A_{\TF}\big(\sum_j\phi_j^{\rm at}\big)^{3/2}\leq K^{1/2}\sum_j\rho_j^{\rm at},\ \ \rho_{\TF}^{5/3}\leq K^{3/2}\sum_j(\rho_j^{\rm at})^{5/3}.
\end{aligned}
\end{equation}
Following \cite[Lemma~6.2]{FNV}, we now estimate $\mathcal F_V(\rho)-\mathcal F_V(\rho_{\TF})$ and apply Lemma~\ref{lem:tf-convexity} to compare the screened potentials $\Phi_r^I$ and $\Phi_r^{I,\TF}$ defined in \eqref{eq:screen}.

\begin{lemma}\label{lem:initial}
Assume that $N\geq Z\geq1$. For every nonempty set $I$ and every radius $r>0$, we have
\begin{equation}\label{eq:2.59}
\|\Phi_r^I-\Phi_r^{I,\TF}\|_{L^\infty(\partial\Omega_r^I)} \leq CZ^{4/3}r^{1/12}.
\end{equation}
\end{lemma}

\begin{proof}
Following \cite{LiebReview}, set $f:=A_{\TF}\min\{\phi_{\TF},Z^2\}^{3/2}$. It then follows from \eqref{eq:tfe} that $0\leq f\leq\rho_{\TF}$, and hence $\int_{\R^3}f dx\leq Z\leq N$. Note that for $0<\delta<Z$, the function $[\min\{\sqrt{\phi_{\TF}},Z\}-\delta]_+$ is Lipschitz and compactly supported, and equals $Z-\delta$ near each $R_j$. Multiplying \eqref{eq:tfpde} by $[\min\{\sqrt{\phi_{\TF}},Z\}-\delta]_+$ and integrating over $\R^3$, we obtain that
\begin{equation}\label{eq:2.60}
\frac12\int_{\{\delta^2<\phi_{\TF}<Z^2\}}\phi_{\TF}^{-1/2}|\nabla\phi_{\TF}|^2 dx\leq4\pi\sum_{j=1}^KZ_j(Z-\delta)=4\pi Z(Z-\delta).
\end{equation}
Letting $\delta\to0^+$ in \eqref{eq:2.60} then yields that
\begin{equation}\label{eq:2.61}
\int_{\R^3}|\nabla\sqrt f|^2 dx
=\frac{9A_{\TF}}{16}\int_{\{\phi_{\TF}<Z^2\}}\phi_{\TF}^{-1/2}|\nabla\phi_{\TF}|^2 dx\leq CZ^2.
\end{equation}
Thus $\sqrt f\in H^1(\R^3)$. Since $\phi_{\TF}\leq V\leq Z/d$, we have $\{\phi_{\TF}>Z^2\}\subset\{d<Z^{-1}\}$. On each set $\Gamma_j$ in \eqref{eq:2.35}, we have $d=d_j$. We therefore deduce from \eqref{eq:tfe} that
\begin{equation}\label{eq:2.62}
\int_{\R^3} V(\rho_{\TF}-f) dx\leq C Z^{5/2}\sum_{j=1}^K\int_{\{d_j< Z^{-1}\}}d_j^{-5/2} dx\leq C_KZ^2.
\end{equation}
Since $0\leq f\leq\rho_{\TF}$, we have $D(f)\leq D(\rho_{\TF})$ and $\int_{\R^3}f^{5/3} dx\leq\int_{\R^3}\rho_{\TF}^{5/3} dx$. We thus conclude from \eqref{eq:massm}, \eqref{eq:tff}, \eqref{eq:2.61} and \eqref{eq:2.62} that
\begin{equation}\label{eq:2.63}
\begin{aligned}
\mathcal E_V(\rho)\leq\mathcal E_V(f)
\leq&\mathcal F_V(\rho_{\TF})+c^{\mathrm W}\int_{\R^3}|\nabla\sqrt f|^2 dx+\int_{\R^3}V(\rho_{\TF}-f) dx\\
\leq&\mathcal F_V(\rho_{\TF})+C_K Z^2.
\end{aligned}
\end{equation}

By H\"older's inequality, we obtain from \eqref{eq:coarse-bounds} and $Z\geq1$ that
\begin{equation*}
\int_{\R^3}\rho^{4/3} dx\leq N^{1/2}\Big(\int_{\R^3}\rho^{5/3} dx\Big)^{1/2}\leq CZ^{5/3}.
\end{equation*}
Together with \eqref{eq:2.63}, this implies that
\begin{equation}\label{eq:2.64}
\mathcal F_V(\rho)-\mathcal F_V(\rho_{\TF})\leq\mathcal E_V(\rho)-\mathcal F_V(\rho_{\TF})+c^{\mathrm D}\int_{\R^3}\rho^{4/3} dx\leq CZ^2.
\end{equation}
It follows from \eqref{eq:tfe} that
\begin{equation*}
\mathcal F_V(\rho)-\mathcal F_V(\rho_{\TF})=D(\rho-\rho_{\TF})
+c^{\TF}\int_{\R^3}\Big[\rho^{5/3}-\rho_{\TF}^{5/3}-\frac53\rho_{\TF}^{2/3}(\rho-\rho_{\TF})\Big] dx.
\end{equation*}
From \eqref{eq:coarse-bounds}, \eqref{eq:aestimates} and \eqref{eq:tfdcom}, we have $\|\rho+\rho_{\TF}\|_{5/3}\leq CZ^{7/5}$ for $Z\geq1$. Since $D(\rho-\rho_{\TF})\geq0$, we deduce from \eqref{eq:tf-convexity} and \eqref{eq:2.64} that
\begin{equation}\label{eq:indensity}
\|\rho-\rho_{\TF}\|_{5/3}\leq C\|\rho+\rho_{\TF}\|_{5/3}^{1/6}\big[\mathcal F_V(\rho)-\mathcal F_V(\rho_{\TF})\big]^{1/2}\leq CZ^{37/30}.
\end{equation}
For every $x\in\R^3$, we have $\int_{\Omega_r^I}|x-y|^{-5/2} dy\leq\int_{B(x,r)}|x-y|^{-5/2} dy+r^{-5/2}|\Omega_r^I|\leq C_Kr^{1/2}$. By H\"older's inequality, it thus gives from \eqref{eq:indensity}  that for $0<r\leq1$,
\begin{equation}\label{eq:2.49}
\big|[(\ind_{\Omega_r^I}(\rho-\rho_{\TF}))*\coul](x)\big|\leq C_Kr^{1/5}Z^{37/30}\leq C_Kr^{1/12}Z^{4/3}.
\end{equation}
We have for $0\leq f\in L^1(\R^3)\cap L^{5/3}(\R^3)$,
\begin{equation}\label{obs:coulomb-interpolation}
\begin{aligned}
\|f*\coul\|_\infty
&\leq\inf_{T>0}\Big[\|f\|_{5/3}\Big(\int_{|z|\leq T}|z|^{-5/2} dz\Big)^{2/5}+T^{-1}\|f\|_1\Big]\\
&\leq\inf_{T>0}\left[CT^{1/5}\|f\|_{5/3}+T^{-1}\|f\|_1\right]
\leq C\|f\|_{5/3}^{5/6}\|f\|_1^{1/6},
\end{aligned}
\end{equation}
where we have minimized in $T$ in the last inequality.
Since $\|\rho+\rho_{\TF}\|_1=N+Z\leq CZ$, we obtain from \eqref{obs:coulomb-interpolation} that for $r\geq1$,
\begin{equation*}
\|(\rho+\rho_{\TF})*\coul\|_\infty\leq C(Z^{7/5})^{5/6}Z^{1/6}=CZ^{4/3}\leq CZ^{4/3}r^{1/12}.
\end{equation*} 
Together with \eqref{eq:2.49} and \eqref{eq:screen}, this completes the proof of Lemma~\ref{lem:initial}.
\end{proof}

\section{The Exterior Thomas--Fermi Problem}\label{eq:section}

Motivated by \cite[Sections~5--6]{FNV} and \cite[Section~12]{SolovejHF}, in this section we study an exterior TF problem and establish comparisons with the neutral molecular TF minimizer.

Fix a nonempty set $I\subset\{1,\ldots,K\}$. We use the notations in \eqref{eq:actived}--\eqref{eq:screen}. Let $\rho$ be a TFDW minimizer with $\int_{\R^3}\rho dx=N\geq Z$. Define for $r>0$
\begin{equation}\label{eq:M}
M_r:=Z-\int_{\Omega_r^I}\rho dx.
\end{equation}
We consider the auxiliary problem
\begin{equation}\label{eq:exttf}
\begin{aligned}
\inf\Big\{F_r(f):f\geq0,\ f\in L^1(\R^3)\cap L^{5/3}(\R^3),\ \int_{\R^3}f dx\leq M_r\Big\},
\end{aligned}
\end{equation}
where the energy functional is defined by
\begin{equation*}
F_r(f):=c^{\TF}\int_{\R^3}f^{5/3} dx-\int_{A_r^I}\Phi_r^I f dx+D(f).
\end{equation*}
Here $A_r^I$ and $\Phi_r^I$ are defined in \eqref{eq:actived} and \eqref{eq:screen}, respectively. If this problem has a minimizer $q_r$, then we define $\phi_r:=\Phi_r^I-q_r*|\cdot|^{-1}$.

Throughout this section, let $0<r\leq2^{-37/30}$ and $0<\beta\leq1$, with $\beta>0$ sufficiently small in terms of $K>0$. Assume that
\begin{equation}\label{eq:cass}
Z_i\geq r^{-3}\text{ \ for \  }i\in I,\ \ \sum_{j\notin I}Z_j\leq Kr^{-3},
\end{equation}
and
\begin{equation}\label{eq:3.4}
\sup_{\partial\Omega_s^I}|\Phi_s^I-\Phi_s^{I,\TF}|\leq\beta s^{-4}\text{ \ for \  }r/100\leq s\leq r.
\end{equation}
We shall establish the following proposition.
\begin{proposition}\label{eq:main}
The problem \eqref{eq:exttf} has a unique minimizer $q_r$ with Lagrange multiplier zero. We also have
\begin{equation}\label{eq:3.5}
D(\ind_{A_r^I}\rho-q_r)\leq F_r(\ind_{A_r^I}\rho)-F_r(q_r)\leq Cr^{-7+2/5}.
\end{equation}
\end{proposition}

To prove Proposition~\ref{eq:main}, we first estimate the exterior mass of $\rho$ and the integral of $\rho^{5/3}$ over the same region. These two estimates do not require the lower bound $Z_i\geq r^{-3}$ for $i\in I$ in \eqref{eq:cass}.
\begin{lemma}\label{lem:3.2}
We have
\begin{equation}\label{eq:intcom}
\int_{A_u^I}\rho dx\leq Cr^{-3}\text{ \ and \  } \int_{A_u^I}\rho^{5/3} dx\leq Cr^{-7}\text{ \ for \  }r/2\leq u\leq r.
\end{equation}
\end{lemma}

\begin{proof}
It follows from \eqref{eq:acomparison}--\eqref{eq:tfdcom} that for all $r>0$,
\begin{equation}\label{eq:tfexterior}
\begin{aligned}
\int_{A_r^I}\rho_{\TF} dx\leq&C\sum_{i\in I}\int_{d_i>r}d_i^{-6} dx+C\sum_{j\notin I}\int_{\R^3}\rho_j^{\rm at} dx \leq C\big(r^{-3}+\sum_{j\notin I}Z_j\big),\\
\int_{A_r^I}\rho_{\TF}^{5/3} dx\leq&C\sum_{i\in I}\int_{d_i>r}d_i^{-10} dx+C\sum_{j\notin I}\int_{\R^3}(\rho_j^{\rm at})^{5/3} dx\leq C\big(r^{-7}+\sum_{j\notin I}Z_j^{7/3}\big).
\end{aligned}
\end{equation}
We then obtain from \eqref{eq:HLSe}, \eqref{eq:tfexterior} and \eqref{eq:cass} that for $r/100\leq t\leq r$,
\begin{equation}\label{eq:upb11}
\begin{aligned}
\int_{A_t^I}\rho_{\TF} dx\leq Ct^{-3},\ \ \int_{A_t^I}\rho_{\TF}^{5/3} dx+D(\ind_{A_t^I}\rho_{\TF})\leq Ct^{-7}.
\end{aligned}
\end{equation}
This then implies from \eqref{obs:coulomb-interpolation} that for $r/100\leq t\leq r$,
\begin{equation}\label{eq:310}
\|(\ind_{A_t^I}\rho_{\TF})*\coul\|_\infty\leq C(t^{-21/5})^{5/6}(t^{-3})^{1/6}=Ct^{-4}.
\end{equation}
We then derive from \eqref{eq:screen}, \eqref{eq:tfe}, \eqref{eq:acomparison} and \eqref{eq:310} that for $r/100\leq t\leq r$,
\begin{equation}\label{eq:312}
\begin{aligned}
\Phi_t^{I,\TF}-\sum_{j\notin I}\frac{Z_j}{d_j}=&\phi_{\TF}-\sum_{j\notin I}\frac{Z_j}{d_j}+(\ind_{A_t^I}\rho_{\TF})*\coul\\
\leq&\sum_{i\in I}\phi_i^{\rm at}+Ct^{-4}\leq Ct^{-4}\text{ \ on \  }\partial\Omega_t^I,
\end{aligned}
\end{equation}
where we have used $\phi_j^{\rm at}\leq Z_j/d_j$ for $j\notin I$. It follows from \eqref{eq:harmoniccom} that for every $r>0$, 
\begin{equation}\label{eq:harmoniccom2}
|I|r h_I\geq1\text{ \ on \  }\partial\Omega_r^I,\ \  |x|h_I(x)\to1\text{ \ as \  }|x|\to\infty.
\end{equation}
From \eqref{eq:3.4}, \eqref{eq:312} and \eqref{eq:harmoniccom2}, we then have $\Phi_t^I-\sum_{j\notin I}Z_j/d_j\leq Ct^{-3}h_I$ on $\partial\Omega_t^I$ for $0<\beta\leq1$. Applying the maximum principle, we obtain that for $r/100\leq t\leq r$,
\begin{equation}\label{eq:313}
\Phi_t^I\leq Ct^{-3}h_I+\sum_{j\notin I}\frac{Z_j}{d_j}\text{ \ in \  }A_t^I.
\end{equation}
For every $t>0$, it follows from \eqref{eq:screen} that $\Phi_t^I-\Phi_t^{I,\TF}$ is harmonic in $A_t^I$ and tends to zero as $|x|\to\infty$. By \eqref{eq:harmoniccom2}, we have $|I|t h_I\geq1$ on $\partial\Omega_t^I$. Applying the maximum principle on each connected component of $A_t^I$, we obtain that
\begin{equation*}
|\Phi_t^I(x)-\Phi_t^{I,\TF}(x)|\leq |I|t h_I(x)\|\Phi_t^I-\Phi_t^{I,\TF}\|_{L^\infty(\partial\Omega_t^I)}\text{ \ for \ }x\in A_t^I.
\end{equation*}
This then implies from \eqref{eq:harmoniccom2} that
\begin{equation}\label{eq:screenm}
\Big|\int_{\Omega_t^I}(\rho-\rho_{\TF}) dx\Big|
=\lim_{|x|\to\infty}|x|\big|\Phi_t^I(x)-\Phi_t^{I,\TF}(x)\big|
\leq |I|t\|\Phi_t^I-\Phi_t^{I,\TF}\|_{L^\infty(\partial\Omega_t^I)}.
\end{equation}
By the maximum principle, \eqref{eq:3.4} also holds in $A_t^I$. This then implies from \eqref{eq:screenm} that for $r/100\leq t\leq r$,
\begin{equation}\label{eq:1harmonic}
|\Phi_t^I-\Phi_t^{I,\TF}|\leq\beta t^{-4}\text{ \ in \ }A_t^I,\ \ \Big|\int_{\Omega_t^I}(\rho-\rho_{\TF}) dx\Big|\leq K\beta t^{-3}
\end{equation}
Set $s:=r/3$, so that $r/100<s<r/2\leq r$.  We then deduce from \eqref{eq:upb11} and \eqref{eq:1harmonic}, that
\begin{equation}\label{eq:annular-mass}
\begin{aligned}
\int_{A_s^I\setminus A_{3s/2}^I}\rho dx=&\int_{\Omega_{3s/2}^I\setminus\Omega_s^I}\rho_{\TF} dx+\int_{\Omega_{3s/2}^I}(\rho-\rho_{\TF}) dx-\int_{\Omega_s^I}(\rho-\rho_{\TF}) dx\\
\leq&\int_{A_s^I}\rho_{\TF} dx+K\beta\big((3s/2)^{-3}+s^{-3}\big)\leq Cr^{-3}.
\end{aligned}
\end{equation}
Applying Lemma~\ref{lem:exterior2} with $Q=Cs^{-3}$ and $t=s$, we obtain from \eqref{eq:cass}, \eqref{eq:313} and \eqref{eq:annular-mass} that
\begin{equation}\label{eq:l1-input}
\int_{A_s^I}\rho dx\leq Cr^{-3}+Cr^{6/5}\Big(\int_{A_{r/2}^I}\rho^{5/3} dx\Big)^{3/5}.
\end{equation}

Let $\chi,\eta\in C^\infty(\R^3;[0,1])$ satisfy that $\chi^2+\eta^2\equiv1$ in $\R^3$, $\eta\equiv0$ on $\Omega_s^I$, $\eta\equiv1$ on $A_{3s/2}^I$ and $|\nabla\chi|^2+|\nabla\eta|^2\leq Cs^{-2}\ind_{A_s^I\setminus A_{3s/2}^I}$ in $\R^3$. Set $f:=\eta^2\rho$. By \eqref{eq:massm}, we have $\mathcal E_V(\rho)\leq\mathcal E_V(\chi^2\rho)$. This then yields from \eqref{eq:binding} that
\begin{equation}\label{eq:318}
c^{\TF}\int_{\R^3}f^{5/3} dx-\int_{\R^3}\Phi_s^I f dx+D(f)-c^{\mathrm D}\int_{\R^3}f^{4/3} dx\leq C(1+s^{-2})\int_{A_s^I\setminus A_{3s/2}^I}\rho dx.
\end{equation}
Since $f\equiv0$ in $\Omega_s^I$ and $c^{\mathrm D}z^{4/3}\leq(c^{\TF}/4)z^{5/3}+Cz$ for $z\geq0$, we derive from \eqref{eq:2.43}, \eqref{eq:cass} and \eqref{eq:313} that
\begin{equation*}
\begin{aligned}
&c^{\TF}\int_{\R^3}f^{5/3} dx+D(f)-\int_{\R^3}\Phi_s^I f dx-c^{\mathrm D}\int_{\R^3}f^{4/3} dx\\
\geq&\frac{c^{\TF}}4\int_{\R^3}f^{5/3} dx+\frac12D(f)-C\int_{\R^3}f dx\\
&+\Big[\frac{c^{\TF}}2\int_{\R^3}f^{5/3} dx+\frac12D(f)-\int_{\R^3}\Big(Cs^{-3}h_I+\sum_{j\notin I}\frac{Z_j}{d_j}\Big)f dx\Big]\\
\geq&\frac{c^{\TF}}4\int_{\R^3}f^{5/3} dx-Cr^{-7}-C\int_{A_s^I}\rho dx,
\end{aligned}
\end{equation*}
We thus obtain from \eqref{eq:annular-mass} and \eqref{eq:318} that
\begin{equation*}
\int_{A_{r/2}^I}\rho^{5/3} dx\leq\int_{\R^3}f^{5/3} dx
\leq C\Big[(1+s^{-2})\int_{A_s^I\setminus A_{3s/2}^I}\rho dx+\int_{A_s^I}\rho dx+r^{-7}\Big].
\end{equation*}
Since $s=r/3$ and $0<r<1$, we deduce from \eqref{eq:annular-mass} that
\begin{equation}\label{eq:preliminary-b}
\int_{A_{r/2}^I}\rho^{5/3} dx\leq C\Big(r^{-7}+\int_{A_s^I}\rho dx\Big).
\end{equation}
Substituting \eqref{eq:preliminary-b} into \eqref{eq:l1-input} yields that
\begin{equation*}
\begin{aligned}
\int_{A_s^I}\rho dx\leq Cr^{-3}+Cr^{6/5}\Big(r^{-7}+\int_{A_s^I}\rho dx\Big)^{3/5}\leq Cr^{-3}+\frac12\int_{A_s^I}\rho dx+Cr^3,
\end{aligned}
\end{equation*}
where we have used Young's inequality in the last inequality. We therefore conclude from \eqref{eq:preliminary-b} that
\begin{equation*}
\int_{A_{r/3}^I}\rho dx\leq Cr^{-3},\ \  \int_{A_{r/2}^I}\rho^{5/3} dx\leq Cr^{-7}.
\end{equation*}
Since $A_u^I\subset A_{r/2}^I\subset A_{r/3}^I$ for $r/2\leq u\leq r$, this proves \eqref{eq:intcom} and completes the proof of Lemma~\ref{lem:3.2}.
\end{proof}

We next prove that \eqref{eq:exttf} has a unique minimizer and establish a lower bound for $\phi_r$ on $\partial\Omega_r^I$.

\begin{lemma}\label{eq:auxiliary}
The problem \eqref{eq:exttf} has a unique minimizer $q_r$ with a Lagrange multiplier $\nu_r\geq0$. Moreover, we have
\begin{equation}\label{eq:positive-boundary}
\phi_r\geq\tfrac45\phi_{\TF}\text{ \ on \ }\partial\Omega_r^I\setminus\{R_j:j\notin I\}.
\end{equation}
\end{lemma}

\begin{proof}
For every $x\in\partial\Omega_r^I\setminus\{R_j:j\notin I\}$, there exists $i_0\in I$ such that $d_{i_0}(x)=r$. Since $Z_{i_0}d_{i_0}(x)^3=Z_{i_0}r^3\geq1$ by \eqref{eq:cass}, we deduce from \eqref{eq:acomparison} and \eqref{eq:aestimates} that
\begin{equation}\label{eq:tf-boundary-lower}
\phi_{\TF}(x)\geq\phi_{i_0}^{\rm at}(x)\geq c_0r^{-4}.
\end{equation}
Take $0<\beta\leq\min\{1,c_0/10\}$. By \eqref{eq:screen} and \eqref{eq:tfe}, we have $\Phi_r^{I,\TF}=\phi_{\TF}+(\ind_{A_r^I}\rho_{\TF})*\coul\geq\phi_{\TF}$ away from the nuclei. From \eqref{eq:3.4}, \eqref{eq:tf-boundary-lower} and $h_I\leq r^{-1}$ on $\partial\Omega_r^I$, we obtain that $\Phi_r^I\geq(c_0-\beta)r^{-4}\geq(c_0/2)r^{-3}h_I$ on $\partial\Omega_r^I$ away from the nuclei. Since $\Phi_r^I-(c_0/2)r^{-3}h_I$ is harmonic in $A_r^I\setminus\{R_j:j\notin I\}$, tends to $+\infty$ as $x\to R_j$ for every $j\notin I$ such that $R_j\in\overline{A_r^I}$ and tends to zero as $|x|\to\infty$, by the minimum principle and \eqref{eq:313} we have
\begin{equation}\label{eq:screened-mass-comparison}
\tfrac12c_0r^{-3}h_I\leq\Phi_r^I\leq Cr^{-3}h_I+\sum_{j\notin I}Z_jd_j^{-1}\text{ \ in \ }A_r^I\setminus\{R_j:j\notin I\}.
\end{equation}
Since $\ind_{\Omega_r^I}\rho$ has bounded support, it gives from \eqref{eq:M} and \eqref{eq:screen} that $|x|\Phi_r^I(x)\to M_r$ as $|x|\to\infty$. Multiplying \eqref{eq:screened-mass-comparison} by $|x|$ and letting $|x|\to\infty$, we thus deduce from \eqref{eq:cass} and \eqref{eq:harmoniccom2} that
\begin{equation}\label{eq:auxiliary-mass}
\tfrac12c_0r^{-3}\leq M_r\leq Cr^{-3},\ \  \int_{A_r^I}\rho dx-M_r=N-Z\geq0,
\end{equation}
where the identity follows from \eqref{eq:M}.

Since $M_r>0$ and $\ind_{A_r^I}\Phi_r^I\in L^{5/2}(\R^3)+L^\infty(\R^3)$ and $\ind_{A_r^I}(x)\Phi_r^I(x)\to0$ as $|x|\to\infty$, we obtain from \cite[Theorem~4.1(i)]{FNV} that the problem \eqref{eq:exttf} has a unique minimizer $q_r$ with a Lagrange multiplier $\nu_r\geq0$ satisfying
\begin{equation}\label{eq:kkt}
\frac{5c^{\TF}}{3}q_r^{2/3}=[\ind_{A_r^I}\Phi_r^I-q_r*|\cdot|^{-1}-\nu_r]_+\text{ \ in \  }\mathbb R^3,\ \ 
\nu_r\Big(M_r-\int_{\R^3}q_r dx\Big)=0.
\end{equation}
This then yields that
\begin{equation}\label{3.335}
q_r\equiv0\text{ \ in \  } \Omega_r^I,\ \ \frac{5c^{\TF}}{3}q_r^{2/3}=[\phi_r-\nu_r]_+\text{ \ in \  }A_r^I.
\end{equation}

It follows from \eqref{eq:harmoniccom} and \eqref{eq:cass} that the potential $Cr^{-3}h_I+\sum_{j\notin I}Z_jd_j^{-1}$ has total charge at most $(C+K)r^{-3}$. From \eqref{eq:313}, we have for every nonnegative $f\in L^1(\R^3)\cap L^{5/3}(\R^3)$,
\begin{equation*}
F_r(f)\geq\frac{c^{\TF}}2\int_{\R^3}f^{5/3} dx+\frac12D(f)+\Big[\frac{c^{\TF}}2\int_{\R^3}f^{5/3} dx+\frac12D(f)-\int_{\R^3}\Big(Cr^{-3}h_I+\sum_{j\notin I}\frac{Z_j}{d_j}\Big)f dx\Big].
\end{equation*}
We then obtain from \eqref{eq:2.43} that
\begin{equation}\label{eq:lowerf}
F_r(f)\geq\frac{c^{\TF}}2\int_{\R^3}f^{5/3} dx+\frac12D(f)-Cr^{-7}.
\end{equation}
Since $F_r(q_r)\leq F_r(0)=0$, it follows from \eqref{eq:upb11}, \eqref{eq:auxiliary-mass} and \eqref{eq:lowerf} that
\begin{equation}\label{eq:tfbounds}
\begin{aligned}
&\int_{\R^3}q_r dx+\int_{A_r^I}\rho_{\TF} dx\leq Cr^{-3},\\
&\int_{\R^3}q_r^{5/3} dx+\int_{A_r^I}\rho_{\TF}^{5/3} dx+D(q_r)+D(\ind_{A_r^I}\rho_{\TF})\leq Cr^{-7}.
\end{aligned}
\end{equation}

Since $q_r-\ind_{A_r^I}\rho_{\TF}\in L^1(\R^3)\cap L^{5/3}(\R^3)$, we have $(q_r-\ind_{A_r^I}\rho_{\TF})*\coul\in C(\R^3)$ and $[(q_r-\ind_{A_r^I}\rho_{\TF})*\coul](x)\to0$ as $|x|\to\infty$. It follows from \eqref{eq:screen} and \eqref{eq:tfe} that
\begin{equation}\label{eq:auxiliary-potential-difference}
\phi_r-\phi_{\TF}\equiv\Phi_r^I-\Phi_r^{I,\TF}-(q_r-\ind_{A_r^I}\rho_{\TF})*\coul\text{ \ in \ }A_r^I,
\end{equation}
where the difference extends continuously to each $R_j\in\overline{A_r^I}$ with $j\notin I$. By the maximum principle, it gives from \eqref{eq:3.4} that $|\Phi_r^I-\Phi_r^{I,\TF}|\leq\beta r^{-4}$ in $A_r^I$. Hence $\phi_r\leq\phi_{\TF}$ almost everywhere in $A_r^I\cap\{(q_r-\ind_{A_r^I}\rho_{\TF})*\coul>\beta r^{-4}\}$. Since $\nu_r\geq0$, we deduce from \eqref{eq:tfe} and \eqref{3.335} that
\begin{equation*}
q_r\leq\ind_{A_r^I}\rho_{\TF}\text{ \ a.e.\ in \ }\{(q_r-\ind_{A_r^I}\rho_{\TF})*\coul>\beta r^{-4}\}.
\end{equation*}
It follows that $\Delta[(q_r-\ind_{A_r^I}\rho_{\TF})*\coul]=4\pi(\ind_{A_r^I}\rho_{\TF}-q_r)\geq0$ in the sense of distributions in $\{(q_r-\ind_{A_r^I}\rho_{\TF})*\coul>\beta r^{-4}\}$. By the weak maximum principle, we thus obtain that
\begin{equation}\label{eq:auxiliary-potential-upper}
(q_r-\ind_{A_r^I}\rho_{\TF})*\coul\leq\beta r^{-4}\text{ \ in \ }\R^3.
\end{equation}
From \eqref{eq:3.4}, \eqref{eq:auxiliary-potential-difference} and \eqref{eq:auxiliary-potential-upper}, we have $\phi_r\geq\phi_{\TF}-2\beta r^{-4}$ on $\partial\Omega_r^I\setminus\{R_j:j\notin I\}$. Since $2\beta\leq c_0/5$, it yields from \eqref{eq:tf-boundary-lower} that $\phi_r\geq(4/5)\phi_{\TF}$ on $\partial\Omega_r^I\setminus\{R_j:j\notin I\}$. This proves \eqref{eq:positive-boundary}. The proof of Lemma~\ref{eq:auxiliary} is thus complete.
\end{proof}

Applying the boundary lower bound in Lemma~\ref{eq:auxiliary}, we prove that the Lagrange multiplier $\nu_r$ is zero and estimate $\phi_r-\phi_{\TF}$.

\begin{lemma}\label{eq:neutral}
The Lagrange multiplier $\nu_r$ in Lemma~\ref{eq:auxiliary} is zero. For $x\in A_r^I$ away from the nuclei, we have $\phi_r(x)\geq(4/5)\phi_{\TF}(x)>0$. Moreover,
\begin{equation}\label{eq:tf-close}
\sup_{A_r^I}|\phi_r-\phi_{\TF}|\leq2\beta r^{-4}.
\end{equation}
\end{lemma}

\begin{proof}
From \eqref{eq:kkt}, we derive the equation
\begin{equation}\label{eq:eequation}
-\Delta\phi_r+k_{\TF}[\phi_r-\nu_r]_+^{3/2} =4\pi\sum_{\substack{j\notin I \\
R_j\in A_r^I}}Z_j\delta_{R_j}\text{ \ in \  }\mathcal{D}'(A_r^I).
\end{equation}
On the set where $(4/5)\phi_{\TF}>\phi_r$ in $A_r^I$, we deduce from \eqref{eq:tfpde}, \eqref{eq:eequation} and $\nu_r\geq0$ that
\begin{equation*}
\Delta\big(\tfrac45\phi_{\TF}-\phi_r\big)=k_{\TF}\big(\tfrac45\phi_{\TF}^{3/2}-[\phi_r-\nu_r]_+^{3/2}\big)\geq k_{\TF}\big(\tfrac45-(\tfrac45)^{3/2}\big)\phi_{\TF}^{3/2}\geq0.
\end{equation*}
By \eqref{eq:tfe} and \eqref{eq:auxiliary-potential-difference}, we have $(4/5)\phi_{\TF}(x)-\phi_r(x)\to-\infty$ as $x\to R_j$ for every $j\notin I$ such that $R_j\in\overline{A_r^I}$. Since $(4/5)\phi_{\TF},\phi_r(x)\to0$ as $|x|\to\infty$, by the maximum principle we obtain from \eqref{eq:positive-boundary} that
\begin{equation}\label{eq:exterior-relative-lower}
\phi_r\geq\tfrac45\phi_{\TF}>0\text{ \ in \ }A_r^I\setminus\{R_j:j\notin I\}.
\end{equation}

If $\nu_r>0$, then we have $\int_{\R^3}q_r dx=M_r$ by \eqref{eq:kkt}. Since $\Phi_r^I(x)\to0$ as $|x|\to\infty$, we also obtain from \eqref{eq:kkt} that $q_r$ is compactly supported. Choose $T>0$ so that $B(0,T)$ contains all nuclei, $\Omega_r^I$ and $\operatorname{supp}q_r$. By Newton's theorem, we obtain from \eqref{eq:M} that
\begin{equation*}
\frac1{4\pi}\int_{\mathbb S^2}\phi_r(T\omega) d\omega=\frac1T\Big(Z-\int_{\Omega_r^I}\rho dx-\int_{\R^3}q_r dx\Big)=0,
\end{equation*}
which contradicts \eqref{eq:exterior-relative-lower}. Thus $\nu_r=0$, and we deduce from \eqref{3.335} and \eqref{eq:exterior-relative-lower} that
\begin{equation}\label{eq:zeromultiplier}
q_r\equiv A_{\TF}\phi_r^{3/2}\text{ \ in \  } A_r^I.
\end{equation}
From \eqref{eq:1harmonic} and \eqref{eq:auxiliary-potential-difference}, we have $\phi_r\geq\phi_{\TF}$ a.e. in $A_r^I\cap\{(q_r-\ind_{A_r^I}\rho_{\TF})*\coul<-\beta r^{-4}\}$. This then implies from \eqref{eq:tfe}, \eqref{3.335} and \eqref{eq:zeromultiplier} that
\begin{equation*}
\Delta[(q_r-\ind_{A_r^I}\rho_{\TF})*\coul]=4\pi(\ind_{A_r^I}\rho_{\TF}-q_r)\leq0
\end{equation*}
in the sense of distributions in $\{(q_r-\ind_{A_r^I}\rho_{\TF})*\coul<-\beta r^{-4}\}$. Since $[(q_r-\ind_{A_r^I}\rho_{\TF})*\coul](x)\to0$ as $|x|\to\infty$, we have $(q_r-\ind_{A_r^I}\rho_{\TF})*\coul\geq-\beta r^{-4}$ in $\R^3$. Together with \eqref{eq:1harmonic}, \eqref{eq:auxiliary-potential-difference} and \eqref{eq:auxiliary-potential-upper}, this proves \eqref{eq:tf-close}. We thus complete the proof of Lemma~\ref{eq:neutral}.
\end{proof}

Following the localization and trial density arguments in \cite[Lemmas~5.2 and 6.9]{FNV}, we next prove the following lemma.

\begin{lemma}\label{eq:localization}
For every $0<\lambda\leq1/2$, we have
\begin{equation}\label{eq:339}
D(\ind_{A_r^I}\rho-q_r)\leq F_r(\ind_{A_r^I}\rho)-F_r(q_r)\leq C(\lambda^{-2}r^{-5}+\lambda^{1/2}r^{-7}).
 \end{equation}
\end{lemma}

\begin{proof}
Let $S_\lambda:=\{x\in\R^3:(1-\lambda)r\leq d_I(x)\leq(1+\lambda)r\}$. Take a Lipschitz cutoff $0\leq\eta_-\leq1$ such that
\begin{equation}\label{eq:3cutoff}
\begin{aligned}
&\eta_-\equiv1\text{ \ on \ }\Omega_{(1-\lambda)r}^I,\ \  \eta_-\equiv0\text{ \ on \ }A_r^I,\\
&|\nabla\eta_-|^2+\big|\nabla\sqrt{1-\eta_-^2}\big|^2\leq C(\lambda r)^{-2}\ind_{S_\lambda}\text{ \ a.e. in \ }\R^3.
\end{aligned}
\end{equation}
Let $f\geq0$ be supported in $A_r^I$ satisfying $\sqrt f\in H^1(\R^3)$ and $\int_{\R^3}f dx\leq\int_{A_r^I}\rho dx$. Since $\eta_-^2\rho\leq\ind_{\Omega_r^I}\rho$ in $\R^3$, we have
\begin{equation*}
\begin{aligned}
-\int_{\R^3}Vf dx+\iint_{\R^3\times\R^3}\frac{f(x)\eta_-^2(y)\rho(y)}{|x-y|} dx dy\leq&-\int_{\R^3}Vf dx+\int_{\R^3}f\bigl[(\ind_{\Omega_r^I}\rho)*|\cdot|^{-1}\bigr] dx\\
=&-\int_{\R^3}\Phi_r^If dx.
\end{aligned}
\end{equation*}
Moreover, we have
\begin{equation*}
\begin{aligned}
\int_{\R^3}(\eta_-^2\rho+f) dx\leq&\int_{\Omega_r^I}\rho dx+\int_{A_r^I}\rho dx=N\text{ \ and \  }\sqrt{\eta_-^2\rho+f}=\sqrt{\eta_-^2\rho}+\sqrt f\in H^1(\R^3).
\end{aligned}
\end{equation*}
Define
\begin{equation}\label{eq:Arf}
\mathcal A_r(f):=c^{\mathrm W}\int_{\R^3}|\nabla\sqrt f|^2 dx+F_r(f).
\end{equation}
We then obtain from \eqref{eq:massm}  that
\begin{equation}\label{eq:3cutupper}
\begin{aligned}
\mathcal{E}_V(\rho) \leq&\mathcal{E}_V(\eta_-^2\rho+f)\\
=&\mathcal{E}_V(\eta_-^2\rho)+\mathcal{E}_V(f)+\iint_{\R^3\times\R^3}\frac{\eta_-^2(x)\rho(x)f(y)}{|x-y|} dx dy\\
\leq&\mathcal{E}_V(\eta_-^2\rho)+\mathcal{A}_r(f)-c^{\mathrm D}\int_{\R^3}f^{4/3} dx \leq\mathcal{E}_V(\eta_-^2\rho)+\mathcal{A}_r(f),
\end{aligned}
\end{equation}

It follows from \eqref{eq:screen} and \eqref{eq:3cutoff} that
\begin{equation}\label{eq:342}
\begin{aligned}
&D(\rho)-D(\eta_-^2\rho)-D(\ind_{A_r^I}\rho)-\int_{\R^3}V(1-\eta_-^2)\rho dx+\int_{A_r^I}\Phi_r^I\rho dx\\
=&D\big((\ind_{\Omega_r^I}-\eta_-^2)\rho\big)
-\int_{S_\lambda\cap\Omega_r^I}\big[V-(\eta_-^2\rho)*\coul\big](1-\eta_-^2)\rho dx\\
\geq&-\int_{S_\lambda\cap\Omega_r^I}\Phi_{(1-\lambda)r}^I(1-\eta_-^2)\rho dx,
\end{aligned}
\end{equation}
where the last inequality follows from $\eta_-^2\rho\geq\ind_{\Omega_{(1-\lambda)r}^I}\rho$ by \eqref{eq:3cutoff}. For $P(z):=c^{\TF}z^{5/3}-c^{\mathrm D}z^{4/3}$, we have $P'(z)\geq cz^{2/3}-C$ for $z\geq0$. Integrating this bound from $\eta_-^2\rho$ to $\rho$, we obtain that for $U\in\R$,
\begin{equation}\label{eq:343}
\begin{aligned}
P(\rho)-P(\eta_-^2\rho)-U(1-\eta_-^2)\rho\geq& c\big[(1-\eta_-^2)\rho\big]^{5/3}-C(1-\eta_-^2)\rho-[U]_+(1-\eta_-^2)\rho\\
\geq&-C(1-\eta_-^2)\rho-C[U]_+^{5/2}.
\end{aligned}
\end{equation}
By the IMS formula, we obtain from \eqref{eq:3cutoff} that
\begin{equation*}
\int_{\R^3}|\nabla\sqrt\rho|^2 dx
\geq\int_{\R^3}|\nabla(\eta_-\sqrt\rho)|^2 dx-C(\lambda r)^{-2}\int_{S_\lambda}\rho dx.
\end{equation*}
Applying \eqref{eq:343} with $U=\Phi_{(1-\lambda)r}^I$ on $S_\lambda\cap\Omega_r^I$, we then obtain from \eqref{eq:342} that
\begin{equation}\label{eq:extlower}
\begin{aligned}
\mathcal E_V(\rho)\geq{}&\mathcal E_V(\eta_-^2\rho)+F_r(\ind_{A_r^I}\rho)-c^{\mathrm D}\int_{A_r^I}\rho^{4/3} dx\\
&-C(1+(\lambda r)^{-2})\int_{S_\lambda}\rho dx-C\int_{S_\lambda}[\Phi_{(1-\lambda)r}^I]_+^{5/2} dx.
\end{aligned}
\end{equation}
Since $r/2\leq(1-\lambda)r\leq r$, it gives from \eqref{eq:intcom} that
\begin{equation}\label{eq:3.46}
\begin{aligned}
&\int_{A_r^I}\rho^{4/3} dx
\leq\Big(\int_{A_r^I}\rho dx\Big)^{1/2}\Big(\int_{A_r^I}\rho^{5/3} dx\Big)^{1/2}\leq Cr^{-5},\\
&(1+(\lambda r)^{-2})\int_{S_\lambda}\rho dx
\leq C(1+(\lambda r)^{-2})r^{-3}\leq C\lambda^{-2}r^{-5}.
\end{aligned}
\end{equation}

It yields from the definition of $S_\lambda$ that
\begin{equation}\label{eq:shellvolume}
S_\lambda\subset\bigcup_{i\in I}\{x\in\R^3:(1-\lambda)r\leq d_i(x)\leq(1+\lambda)r\},\ \  |S_\lambda|\leq C_K\lambda r^3.
\end{equation}
We have
\begin{equation}\label{eq:349}
\Big(\sum_{j\notin I}\frac{Z_j}{d_j}\Big)^a\leq\Big(\sum_{j\notin I}Z_j\Big)^a\sum_{j\notin I}d_j^{-a}\text{ \ for \  }a>1,
\end{equation}
Since $r/2\leq(1-\lambda)r\leq d_I$ on $S_\lambda$, we derive from \eqref{eq:313} that
\begin{equation}\label{eq:350}
\Phi_{(1-\lambda)r}^I\leq Cr^{-4}+\sum_{j\notin I}\frac{Z_j}{d_j}\text{ \ on \  }S_\lambda.
\end{equation}
Fix $i\in I$ and $j\notin I$, and set $a:=|R_i-R_j|$. We have for $a>0$ and $t>0$ with $t\ne a$,
\begin{equation*}
\int_{\partial B(R_i,t)}d_j^{-5/2} dS\leq\frac{8\pi}{|t-a|^{1/2}},
\end{equation*}
For $a=0$, the integral equals $4\pi t^{-1/2}$. Integrating in $t$ over $[(1-\lambda)r,(1+\lambda)r]$, we then obtain from \eqref{eq:shellvolume} that
\begin{equation}\label{ana}
\int_{S_\lambda}d_j^{-5/2} dx\leq C_K\sup_{a\geq0}\int_{(1-\lambda)r}^{(1+\lambda)r}|t-a|^{-1/2} dt\leq C_K(\lambda r)^{1/2}\text{ \ for \ }j\notin I.
\end{equation}
We thus obtain from \eqref{eq:cass}, \eqref{eq:shellvolume}--\eqref{ana} that
\begin{equation}\label{eq:351}
\begin{aligned}
\int_{S_\lambda}[\Phi_{(1-\lambda)r}^I]_+^{5/2} dx\leq&Cr^{-10}|S_\lambda|+C\Big(\sum_{j\notin I}Z_j\Big)^{5/2}\sum_{j\notin I}\int_{S_\lambda}d_j^{-5/2} dx\\
\leq&C\lambda r^{-7}+C\Big(\sum_{j\notin I}Z_j\Big)^{5/2}(\lambda r)^{1/2}\leq C\lambda^{1/2}r^{-7}.
\end{aligned}
\end{equation}
Consequently, we deduce from \eqref{eq:3cutupper}, \eqref{eq:extlower}, \eqref{eq:3.46} and \eqref{eq:351} that
\begin{equation}\label{eq:three-cut}
F_r(\ind_{A_r^I}\rho)\leq\mathcal A_r(f)+C(\lambda^{-2}r^{-5}+\lambda^{1/2}r^{-7}).
\end{equation}

Choose a Lipschitz cutoff $0\leq\eta_+\leq1$ such that
\begin{equation}\label{eq:trial-cutoff}
\eta_+\equiv0\text{ \ on \ }\Omega_{(1+\lambda/4)r}^I,\ \ 
\eta_+\equiv1\text{ \ on \ }A_{(1+\lambda)r}^I,\ \ 
|\nabla\eta_+|\leq C(\lambda r)^{-1}\text{ \ a.e. in \ }\R^3.
\end{equation}
Set $E_{\lambda r}:=\bigcup_{j\notin I}B(R_j,\lambda r)$. Take $\theta\in C^\infty(\R^3;[0,1])$ such that
\begin{equation*}
\theta\equiv0\text{ \ on \ }\bigcup_{j\notin I}B(R_j,\lambda r/2),\ \ 
\theta\equiv1\text{ \ on \ }E_{\lambda r}^c,\ \ 
|\nabla\theta|\leq C(\lambda r)^{-1}\text{ \ in \ }\R^3.
\end{equation*}
If $I=\{1,\cdots,K\}$, then we take $\theta\equiv1$ in $\R^3$. Define $f:=\eta_+^2\theta^2q_r$, so that $0\leq f\leq q_r$ and $\int_{\R^3}f dx\leq M_r\leq\int_{A_r^I}\rho dx$ by \eqref{eq:auxiliary-mass} and \eqref{eq:kkt}. 

By Lemma~\ref{eq:neutral}, we obtain from \eqref{eq:eequation} that $\phi_r>0$ and $\Delta\phi_r=k_{\TF}\phi_r^{3/2}$ in $A_r^I\setminus\{R_j:j\notin I\}$. Let $0\leq\xi_R\leq1$ be a smooth cutoff satisfying $\xi_R\equiv1$ in $B(0,R)$, $\xi_R\equiv0$ on $B^c(0,2R)$ and $|\nabla\xi_R|\leq C/R$. Multiplying the equation by $(\eta_+\theta\xi_R)^2\sqrt{\phi_r}$ and integrating over $\R^3$, we then obtain  that
\begin{equation*}
\int_{A_r^I}(\eta_+\theta\xi_R)^2\phi_r^{-1/2}|\nabla\phi_r|^2 dx\leq16\int_{A_r^I}\phi_r^{3/2}|\nabla(\eta_+\theta\xi_R)|^2 dx.
\end{equation*}
From \eqref{eq:zeromultiplier}, we have $\phi_r^{3/2}=q_r/A_{\TF}\in L^1(A_r^I)$. Since $|\nabla(\eta_+\theta\xi_R)|$ is uniformly bounded for $R\geq1$, letting $R\to\infty$ yields from dominated convergence that
\begin{equation}\label{eq:356}
\int_{A_r^I}\eta_+^2\theta^2\phi_r^{-1/2}|\nabla\phi_r|^2 dx\leq16\int_{A_r^I}\phi_r^{3/2}|\nabla(\eta_+\theta)|^2 dx.
\end{equation}
It follows from \eqref{eq:trial-cutoff} that $|\nabla(\eta_+\theta)|\leq C(\lambda r)^{-1}$. We therefore obtain from \eqref{eq:tfbounds}, \eqref{eq:zeromultiplier} and \eqref{eq:356} that
\begin{equation}\label{eq:357}
\int_{\R^3}|\nabla\sqrt f|^2 dx\leq 20\int_{A_r^I}q_r|\nabla(\eta_+\theta)|^2 dx\leq C(\lambda r)^{-2}\int_{\R^3}q_r dx\leq C\lambda^{-2}r^{-5}.
\end{equation}
Together with $\int_{\R^3}f dx<\infty$, this implies that $\sqrt f\in H^1(\R^3)$.

Since $|E_{\lambda r}|\leq C_K(\lambda r)^3$, the rearrangement inequality (cf. \cite[Section~3.4]{LiebLoss}) gives that $\int_{E_{\lambda r}}d_j^{-5/2} dx\leq C(\lambda r)^{1/2}$ for each $j\notin I$. From \eqref{eq:screen}, we have $\Phi_r^I\leq\Phi_{(1-\lambda)r}^I$. Using $h_I\leq r^{-1}$ in $A_r^I$, we thus obtain from \eqref{eq:cass}, \eqref{eq:313}, \eqref{eq:349} and \eqref{eq:351} that
\begin{equation}\label{eq:trial-potential}
\begin{aligned}
&\int_{(S_\lambda\cup E_{\lambda r})\cap A_r^I}[\Phi_r^I]_+^{5/2} dx\\
\leq&\int_{S_\lambda}[\Phi_{(1-\lambda)r}^I]_+^{5/2} dx
+Cr^{-10}(\lambda r)^3+C\Big(\sum_{j\notin I}Z_j\Big)^{5/2}(\lambda r)^{1/2}
\leq C\lambda^{1/2}r^{-7}.
\end{aligned}
\end{equation}
Moreover, it gives from \eqref{eq:kkt} that $q_r\leq A_{\TF}[\Phi_r^I]_+^{3/2}$ in $A_r^I$. Since $0\leq f\leq q_r$ and $f\equiv q_r$ outside $S_\lambda\cup E_{\lambda r}$, by the minimality of $q_r$ and the bound $D(f)\leq D(q_r)$ we derive from \eqref{eq:trial-potential} that
\begin{equation}\label{eq:trial-energy-cost}
0\leq F_r(f)-F_r(q_r)\leq\int_{A_r^I}\Phi_r^I(q_r-f) dx\leq A_{\TF}\int_{(S_\lambda\cup E_{\lambda r})\cap A_r^I}[\Phi_r^I]_+^{5/2} dx
\leq C\lambda^{1/2}r^{-7}.
\end{equation}
Consequently, we conclude from \eqref{eq:Arf}, \eqref{eq:357} and \eqref{eq:trial-energy-cost} that
\begin{equation}\label{eq:Afcost}
\mathcal A_r(f)-F_r(q_r)\leq C(\lambda^{-2}r^{-5}+\lambda^{1/2}r^{-7}).
\end{equation}

Since $\nu_r=0$, we deduce from \eqref{eq:kkt} that
\begin{equation*}
\frac{5c^{\TF}}{3}q_r^{2/3}+q_r*|\cdot|^{-1}-\ind_{A_r^I}\Phi_r^I\geq0\text{ \ in \  }\mathbb R^3,
\end{equation*}
where the equality holds in $\{q_r>0\}$. This gives that
\begin{equation*}
\int_{\R^3}\Big(\frac{5c^{\TF}}{3}q_r^{2/3}+q_r*\coul-\ind_{A_r^I}\Phi_r^I\Big)(\ind_{A_r^I}\rho-q_r) dx\geq0.
\end{equation*}
We then obtain that
\begin{equation}\label{eq:unmini}
\begin{aligned}
&F_r(\ind_{A_r^I}\rho)-F_r(q_r)\\
\geq& D(\ind_{A_r^I}\rho-q_r)
+c^{\TF}\int_{\R^3}\left[\ind_{A_r^I}\rho^{5/3}-q_r^{5/3}-\frac53q_r^{2/3}(\ind_{A_r^I}\rho-q_r)\right] dx
\geq0 .
\end{aligned}
\end{equation}
 Together with \eqref{eq:three-cut} and \eqref{eq:Afcost}, this proves \eqref{eq:339}. We thus complete the proof of Lemma~\ref{eq:localization}.
\end{proof}

We now complete the proof of Proposition~\ref{eq:main}.

\begin{proof}[Proof of Proposition~\ref{eq:main}]
By Lemmas~\ref{eq:auxiliary} and \ref{eq:neutral}, the problem \eqref{eq:exttf} has a unique minimizer $q_r$ whose Lagrange multiplier is zero. Take $\lambda:=r^{4/5}/2$. Since $0<r\leq2^{-37/30}<1$, we have $0<\lambda\leq1/2$. We deduce from \eqref{eq:339} that
\begin{equation*}
D(\ind_{A_r^I}\rho-q_r)\leq F_r(\ind_{A_r^I}\rho)-F_r(q_r)
\leq C\big(\lambda^{-2}r^{-5}+\lambda^{1/2}r^{-7}\big)\leq Cr^{-7+2/5}.
\end{equation*}
This proves \eqref{eq:3.5} and completes the proof of Proposition~\ref{eq:main}.
\end{proof}

\subsection{Exterior Thomas--Fermi Comparisons}
In this subsection, we adapt the exterior TF comparison in \cite[Lemma~6.8]{FNV} to the molecular case, where the balls $B(R_i,r)$ with $i\in I$ may overlap and some $R_j$ with $j\notin I$ may lie in $A_r^I$. We first prove the following lemma.

\begin{lemma}\label{ana:tf-difference}
There exist constants $L>1$ and $C>0$, depending only on $K$, such that
\begin{equation}\label{ana:eq-global-difference}
|\phi_r(x)-\phi_{\TF}(x)|\leq Cr^{1/4}\sum_{i\in I}d_i(x)^{-17/4}\text{ \ for \  }x\in A_{Lr}^I.
\end{equation}
\end{lemma}

\begin{proof}
By Lemma~\ref{eq:neutral}, we have $\nu_r=0$. We then obtain from \eqref{eq:tfpde} and \eqref{eq:eequation} that
\begin{equation}\label{ana:eq-common-sources}
-\Delta z+k_{\TF}z^{3/2}=4\pi\sum_{\substack{j\notin I\\R_j\in A_r^I}}Z_j\delta_{R_j}\text{ \ in \  }\mathcal D'(A_r^I)\text{ \ for \  }z=\phi_r,\phi_{\TF}.
\end{equation}
The definitions of $\phi_r$ and $\phi_{\TF}$ give that
\begin{equation*}
\phi_r-\phi_{\TF}\equiv(\rho_{\TF}-\ind_{\Omega_r^I}\rho-q_r)*\coul\text{ \ in \  }\R^3.
\end{equation*}
Since $\rho_{\TF}-\ind_{\Omega_r^I}\rho-q_r\in L^1(\R^3)\cap L^{5/3}(\R^3)$, by \cite[Lemma~II.25]{LiebSimon}, we conclude that $\phi_r-\phi_{\TF}$ extends continuously to $\R^3$ and tends to zero as $|x|\to\infty$.
It follows from \eqref{ana:eq-common-sources} that
\begin{equation}\label{ana:eq-linear-difference}
(-\Delta+\mathcal V)(\phi_r-\phi_{\TF})=0\text{ \ in \  }\mathcal D'(A_r^I),
\end{equation}
where $\mathcal V(x):=(3k_{\TF}/2)\int_0^1\sqrt{(1-t)\phi_{\TF}(x)+t\phi_r(x)} dt\geq0$.

For every $i\in I$, we have $Z_i^{-1/3}\leq r$ by \eqref{eq:cass}. By the atomic Sommerfeld bound in \cite[Theorem~4.2]{FNV}, we may therefore choose $L>1$ independently of $i$ and $Z_i$ such that
\begin{equation*}
\phi_i^{\rm at}(x)\geq\frac45\Big(\frac{12}{k_{\TF}}\Big)^2d_i(x)^{-4}\text{ \ whenever \ }d_i(x)\geq Lr.
\end{equation*}
Together with \eqref{eq:acomparison} and \eqref{eq:exterior-relative-lower}, we obtain that for every $i\in I$,
\begin{equation*}
\mathcal V\geq\tfrac{3k_{\TF}}2\sqrt{\tfrac45\phi_{\TF}}\geq\tfrac{3k_{\TF}}2\sqrt{\tfrac45\phi_i^{\rm at}}\geq\tfrac{72}{5}d_i^{-2}\text{ \ a.e. in \ }A_{Lr}^I.
\end{equation*}
This gives that
\begin{equation}\label{ana:eq-linear-supersolution-check}
(-\Delta+\mathcal V)\Big(Cr^{1/4}\sum_{i\in I}d_i^{-17/4}\Big)=Cr^{1/4}\sum_{i\in I}d_i^{-17/4}\Big(\mathcal V-\frac{221}{16}d_i^{-2}\Big)
\geq0\text{ \ in \ }\mathcal D'(A_{Lr}^I).
\end{equation}
Since $\min_{i\in I}d_i(x)=Lr$ on $\partial\Omega_{Lr}^I$, we take $C\geq C_KL^{17/4}$ and deduce from \eqref{eq:tf-close} that
\begin{equation*}
Cr^{1/4}\sum_{i\in I}d_i^{-17/4}\geq CL^{-17/4}r^{-4}\geq C_Kr^{-4}\geq|\phi_r-\phi_{\TF}|\text{ \ on \  }\partial\Omega_{Lr}^I.
\end{equation*}
Since both $\phi_r-\phi_{\TF}$ and $Cr^{1/4}\sum_{i\in I}d_i^{-17/4}$ tend to zero as $|x|\to\infty$, by the comparison principle, we deduce from \eqref{ana:eq-linear-difference} and \eqref{ana:eq-linear-supersolution-check} that \eqref{ana:eq-global-difference} holds. This completes the proof of Lemma~\ref{ana:tf-difference}.
\end{proof}

The pointwise estimate in Lemma~\ref{ana:tf-difference} gives the following Coulomb potential bound of $|q_r-\rho_{\TF}|$.

\begin{lemma}\label{ana:power-tail}
After increasing $L>1$ in Lemma~\ref{ana:tf-difference} if necessary, for every $I\subset I'\subset\{1,\ldots,K\}$, $s\geq Lr>0$ and $x\in\R^3$, we have
\begin{equation}\label{ana:eq-eighth-tail}
\int_{A_s^{I'}}\frac{|q_r(y)-\rho_{\TF}(y)|}{|x-y|} dy\leq C(r/s)^{1/4}s^{-4},
\end{equation}
where $C>0$ depends only on $K$.
\end{lemma}

\begin{proof}
Set $w:=\phi_r-\phi_{\TF}$. We then obtain from \eqref{eq:tfe}, \eqref{eq:zeromultiplier} and \eqref{ana:eq-common-sources} that
\begin{equation*}
\Delta w=4\pi(q_r-\rho_{\TF})\text{ \ in \ }\mathcal D'(A_r^I),\ \  \operatorname{sgn}(w)(q_r-\rho_{\TF})=|q_r-\rho_{\TF}|\text{ \ a.e. in \ }A_r^I.
\end{equation*}
Since $q_r-\rho_{\TF}\in L^{5/3}(A_r^I)$, we have $w\in W^{2,5/3}_{\rm loc}(A_r^I)$. Then using $\Delta(w^2+\delta^2)^{1/2}$ and letting $\delta\to0^+$, we deduce that
\begin{equation}\label{ana:eq-absolute-source}
\Delta|w|\geq4\pi|q_r-\rho_{\TF}|\text{ \ in \ }\mathcal D'(A_r^I).
\end{equation}

Let $L$ be the constant in Lemma~\ref{ana:tf-difference} and take $s\geq2KLr$. Set $R:=s/(2K)$.  Since $q_r,\rho_{\TF}\in L^1(\R^3)\cap L^{5/3}(\R^3)$, $P:=(\ind_{A_s^I}|q_r-\rho_{\TF}|)*\coul$ is continuous in $\R^3$ and tends to zero as $|x|\to\infty$. From \eqref{ana:eq-absolute-source}, we have $\Delta(P+|w|)\geq0$ in $\mathcal D'(A_R^I)$. Since $|I|Rh_I\geq1$ on $\partial\Omega_R^I$ by \eqref{eq:harmoniccom2} and $P$, $w$ and $h_I$ tend to zero as $|x|\to\infty$, the maximum principle gives that
\begin{equation*}
P(x)+|w(x)|\leq\Big(\|P\|_\infty+\sup_{\partial\Omega_R^I}|w|\Big)|I|Rh_I(x)\text{ \ for \ }x\in A_R^I.
\end{equation*}
For $d_I(x)\geq s$, we have $|I|Rh_I(x)=R\sum_{i\in I}d_i(x)^{-1}\leq KR/s=1/2$. Since $P$ is harmonic in the bounded set $\{d_I<s\}$, the maximum principle and $R\geq Lr$ give that
\begin{equation*}
\|P\|_\infty=\sup_{\{d_I\geq s\}}P\leq\tfrac12\Big(\|P\|_\infty+\sup_{\partial\Omega_R^I}|w|\Big),
\end{equation*}
so that
\begin{equation*}
\|P\|_\infty\leq\sup_{\partial\Omega_R^I}|w|\leq Cr^{1/4}R^{-17/4}\leq C(r/s)^{1/4}s^{-4},
\end{equation*}
where we have used \eqref{ana:eq-global-difference} by continuity. Since $A_s^{I'}\subset A_s^I$, this proves \eqref{ana:eq-eighth-tail}. The proof of Lemma~\ref{ana:power-tail} is thus complete.
\end{proof}

\section{Proof of Theorems~\ref{thm:1.1}-- \ref{thmm:ener}}\label{iter:section}

In this section, we complete the proof of Theorems~\ref{thm:1.1}-- \ref{thmm:ener}. Let $\rho$ be a minimizer of \eqref{eq:inf} with mass $N\geq Z$. For every nonempty $I\subset\{1,\ldots,K\}$ and $s>0$, we use $\Omega_s^I$ from \eqref{eq:active-inner} and the screened potentials from \eqref{eq:screen}. Set
\begin{equation}\label{iter:screening-error-definition}
\mathcal H_I(s):=\sup_{\partial\Omega_s^I}|\Phi_s^I-\Phi_s^{I,\TF}|.
\end{equation}
We first  establish a screening bound for every enlarged set of nuclei.

\begin{proposition}\label{iter:enlarged-screening}
There exist constants $0<\beta\leq1$, $L>1$ and $C>0$, depending only on $K$, such that if $I\ne\varnothing$, $0<r\leq2^{-37/30}$ and
\begin{equation*}
\mathcal{H}_I(t)\leq\beta t^{-4}\text{ \ for \  }r/100\leq t\leq r,\ \  \min_{i\in I}Z_i\geq r^{-3},\ \  \sum_{j\notin I}Z_j\leq Kr^{-3},
\end{equation*}
then we have for every $I\subset I'\subset\{1,\ldots,K\}$ and $s\geq Lr$,
\begin{equation}\label{iter:improvement}
\mathcal H_{I'}(s)\leq C\left[\left(\frac rs\right)^{1/4}s^{-4}+r^{-23/6}\right].
\end{equation}
\end{proposition}
\begin{proof}
By \eqref{eq:intcom} and \eqref{eq:tfbounds}, we have $\|\ind_{A_r^I}\rho-q_r\|_1\leq Cr^{-3}$ and $\|\ind_{A_r^I}\rho+q_r\|_{5/3}\leq Cr^{-21/5}$. Applying \eqref{eq:tf-convexity} to $\ind_{A_r^I}\rho$ and $q_r$, we deduce from  \eqref{eq:3.5} and \eqref{eq:unmini} that
\begin{equation}\label{iter:exterior-input}
\|\ind_{A_r^I}\rho-q_r\|_{5/3}^2\leq C\|\ind_{A_r^I}\rho+q_r\|_{5/3}^{1/3}\big(F_r(\ind_{A_r^I}\rho)-F_r(q_r)\big)
\leq C(r^{-21/5})^{1/3}r^{-7+2/5}=Cr^{-8}.
\end{equation}
Since $\Omega_r^I\subset\Omega_s^{I'}$, we obtain from \eqref{eq:tfe}, \eqref{eq:screen} and the definition of $\phi_r$ given below \eqref{eq:exttf} that
\begin{equation}\label{iter:decomposition}
\begin{aligned}
&\Phi_s^{I'}-\Phi_s^{I',\TF}=\big[\ind_{\Omega_s^{I'}}(\rho_{\TF}-\rho)\big]*\coul\\
=&\phi_r-\phi_{\TF}+\bigl[\ind_{A_s^{I'}}(q_r-\rho_{\TF})\bigr]*\coul+\bigl[\ind_{\Omega_s^{I'}}(q_r-\ind_{A_r^I}\rho)\bigr]*\coul\text{ \ on \  }\partial\Omega_s^{I'}.
\end{aligned}
\end{equation}

Choose $L$ as in Lemma~\ref{ana:power-tail}. Since $d_i(x)\geq s$ for $x\in\partial\Omega_s^{I'}$ and $i\in I$, we obtain from \eqref{ana:eq-global-difference} and \eqref{ana:eq-eighth-tail} that
\begin{equation}\label{iter:potential-difference-bound}
\sup_{\partial\Omega_s^{I'}}|\phi_r-\phi_{\TF}|+\big\|[\ind_{A_s^{I'}}(q_r-\rho_{\TF})]*\coul\big\|_\infty\leq C(r/s)^{1/4}s^{-4}.
\end{equation}
For the last term in \eqref{iter:decomposition}, we use $\|\ind_{A_r^I}\rho-q_r\|_1\leq Cr^{-3}$ and \eqref{iter:exterior-input} in \eqref{obs:coulomb-interpolation} to obtain that
\begin{equation}\label{iter:truncated-potential-bound}
\big\|[\ind_{\Omega_s^{I'}}(q_r-\ind_{A_r^I}\rho)]*\coul\big\|_\infty\leq\big\||q_r-\ind_{A_r^I}\rho|*\coul\big\|_\infty\\
\leq C(r^{-4})^{5/6}(r^{-3})^{1/6}=Cr^{-23/6}.
\end{equation}
Substituting \eqref{iter:potential-difference-bound} and \eqref{iter:truncated-potential-bound} into \eqref{iter:decomposition} proves \eqref{iter:improvement}, and hence completes the proof of Proposition~\ref{iter:enlarged-screening}.
\end{proof}

We adapt the iteration in \cite[Lemmas~6.1 and 6.3]{FNV} and \cite[Lemma~10.3]{SolovejHF} by selecting the nuclei according to their charges at each radius. Let $F:=\{1,\ldots,K\}$, $M:=\max_{j\in F}Z_j$ and let $\beta,L,C$ be as in Proposition~\ref{iter:enlarged-screening}.

\begin{lemma}\label{iter:finite-screening}
There exists a constant $0<D_*\leq2^{-37/30}$, depending only on $K$, such that
\begin{equation}\label{iter:full-screening}
\mathcal H_F(s)\leq s^{-4+10^{-5}}\text{ \ for \  }0<s\leq D_*.
\end{equation}
\end{lemma}

\begin{proof}
Set $a:=1.001$ and $\varepsilon:=10^{-5}$. Let $J\subset F$ be nonempty and let $0<t\leq\min\{1,M^{-1/(3a)}\}$. If $Z<1$, we obtain from \eqref{eq:coarse-bounds}, \eqref{eq:aestimates} and \eqref{eq:tfdcom} that $\|\rho+\rho_{\TF}\|_1+\|\rho+\rho_{\TF}\|_{5/3}\leq C$. It then follows from \eqref{eq:screen} and \eqref{obs:coulomb-interpolation} that $\mathcal H_J(t)\leq\|(\rho+\rho_{\TF})*\coul\|_\infty\leq C$. If $Z\geq1$, we use $Z\leq KM$ and $M\leq t^{-3a}$ in \eqref{eq:2.59} to obtain that $\mathcal H_J(t)\leq CM^{4/3}t^{1/12}\leq Ct^{1/12-4a}$. We have in both cases 
\begin{equation}\label{iter:initial-screening} 
\mathcal H_J(t)t^{4-\varepsilon}\leq C_0t^{49/12-\varepsilon-4a}\text{ \ for \ }0<t\leq\min\{1,M^{-1/(3a)}\},
\end{equation}
where $C_0>0$ depends only on $K$.

Since $49/12-\varepsilon-4a>0$, $(a-1)/4-\varepsilon>0$ and $4-\varepsilon-23a/6>0$, we may choose $0<D_*\leq2^{-37/30}$, depending only on $K$, such that
\begin{equation}\label{iter:small-endpoint}
\begin{aligned}
&C_0D_*^{49/12-\varepsilon-4a}\leq1,\ \  D_*^\varepsilon\leq\beta,\ \ 
D_*^{a-1}\leq\min\{1/2,1/L\},\\
&C\Big(D_*^{(a-1)/4-\varepsilon}+D_*^{4-\varepsilon-23a/6}\Big)\leq1.
\end{aligned}
\end{equation}
For $s>0$, set $I(s):=\{i\in F:Z_i\geq s^{-3}\}$. We claim that for $0<s\leq D_*$,
\begin{equation}\label{iter:state-screening}
\mathcal H_J(s)\leq s^{-4+\varepsilon}\text{ \ for every nonempty \  }J\text{ \ with \  }I(s)\subset J\subset F.
\end{equation}
For $0<s\leq\min\{M^{-1/(3a)},D_*\}$, this follows from \eqref{iter:initial-screening} and  the first inequality in \eqref{iter:small-endpoint} for every nonempty $J\subset F$.

Suppose that \eqref{iter:state-screening} holds for all $0<t\leq T$, where $M^{-1/(3a)}\leq T<D_*$. Fix $T<s\leq\min\{2T,D_*\}$ and $I(s)\subset J\subset F$. From \eqref{iter:small-endpoint}, we have $s^a\leq s/2\leq T$, $Ls^a\leq s$ and $s^a\leq2^{-37/30}$. Since $s>M^{-1/(3a)}$, we have $s^{-3a}<M$, so $I(s^a)$ is nonempty. The definition of  $I(s^a)$ gives that
\begin{equation}\label{iter:state-charges}
\min_{i\in I(s^a)}Z_i\geq s^{-3a},\qquad \sum_{j\notin I(s^a)}Z_j\leq Ks^{-3a}.
\end{equation}
For $s^a/100\leq t\leq s^a$, we have $I(t)\subset I(s^a)\subset I(s)\subset J$. Since $t\leq T$, we take the set $I(s^a)$ in \eqref{iter:state-screening} at radius $t$. Together with \eqref{iter:small-endpoint}, this gives that
\begin{equation}\label{iter:step-input}
\mathcal H_{I(s^a)}(t)\leq t^{-4+\varepsilon}\leq\beta t^{-4}\text{ \ for \ }s^a/100\leq t\leq s^a.
\end{equation}
By Proposition~\ref{iter:enlarged-screening} at radius $s^a$ with $I=I(s^a)$ and $I'=J$, we deduce from \eqref{iter:improvement}, \eqref{iter:state-charges} and \eqref{iter:step-input} that
\begin{equation*}
s^{4-\varepsilon}\mathcal H_J(s)
\leq C\left[(s^a/s)^{1/4}s^{-\varepsilon}+s^{-23a/6}s^{4-\varepsilon}\right]=C\left[s^{(a-1)/4-\varepsilon}+s^{4-\varepsilon-23a/6}\right]\leq1,
\end{equation*}
where we have used $s\leq D_*$ and \eqref{iter:small-endpoint} in the last inequality. Thus \eqref{iter:state-screening} extends from $(0,T]$ to $(0,\min\{2T,D_*\}]$. Starting with $T=\min\{M^{-1/(3a)},D_*\}>0$, we obtain \eqref{iter:state-screening} on $(0,D_*]$ after finitely many iterations. Taking $J=F$ in \eqref{iter:state-screening} proves \eqref{iter:full-screening}. This completes the proof of Lemma~\ref{iter:finite-screening}.
\end{proof}

We are now ready to prove Theorem~\ref{thm:1.1}.

\begin{proof}[Proof of Theorem~\ref{thm:1.1}]
It suffices to consider the case $N\geq Z$. Let $D_*$ be as in Lemma~\ref{iter:finite-screening}. From \eqref{eq:screen1}, \eqref{iter:screening-error-definition}, \eqref{iter:full-screening} and \eqref{iter:small-endpoint}, we obtain that for $0<t\leq D_*$,
\begin{equation}\label{obs:small-screening}
\sup_{\partial\Omega_t}|\Phi_t-\Phi_t^{\TF}|=\mathcal H_F(t)\leq t^{-4+10^{-5}}\leq D_*^{10^{-5}}t^{-4}\leq\beta t^{-4}.
\end{equation}
Thus \eqref{eq:3.4} holds with $I=F$ for every $0<r\leq D_*$. Since $\sum_{j\notin F}Z_j=0$, we apply Lemma~\ref{lem:3.2} with $I=F$ and $u=r=D_*$ to deduce that $\int_{A_{D_*}}\rho dx\leq CD_*^{-3}$. Since $\int_{\R^3}\rho dx=N$ and $\int_{\R^3}\rho_{\TF} dx=Z$, it follows from \eqref{eq:screenm} and \eqref{obs:small-screening} that
\begin{equation*}
N-Z=\int_{\Omega_{D_*}}(\rho-\rho_{\TF}) dx+\int_{A_{D_*}}(\rho-\rho_{\TF}) dx
\leq KD_*\mathcal H_F(D_*)+\int_{A_{D_*}}\rho dx\leq C_KD_*^{-3}.
\end{equation*}
Since $D_*>0$ depends only on $K>0$, this completes the proof of Theorem~\ref{thm:1.1}.
\end{proof}

We next use \eqref{obs:coulomb-interpolation} and \eqref{obs:small-screening} to prove Theorem~\ref{thm:screen}.

\begin{proof}[Proof of Theorem~\ref{thm:screen}]
From \eqref{eq:tfexterior} with $I=F$, we have
\begin{equation}\label{obs:tf-upper-tail}
\int_{A_r}\rho_{\TF} dx\leq Cr^{-3},\ \  \int_{A_r}\rho_{\TF}^{5/3} dx\leq Cr^{-7}\text{ \ for \  }r>0.
\end{equation}
Moreover, it follows from \eqref{eq:screenm} and \eqref{obs:small-screening} that for $0<r\leq D_*$,
\begin{equation}\label{obs:core-mass}
\Big|\int_{\Omega_r}(\rho-\rho_{\TF}) dx\Big|\leq Cr^{-3+10^{-5}}.
\end{equation}
Define
\begin{equation}\label{obs:harmonic-weight}
h(x):=K^{-1}\sum_{j=1}^K|x-R_j|^{-1}.
\end{equation}
For $0<r\leq D_*$, we apply \eqref{eq:313} with $I=F$ and $t=r$, using \eqref{obs:small-screening} and $\sum_{j\notin F}Z_j=0$. Since $h_F\equiv h$ in $A_r$ by \eqref{eq:harmoniccom} and \eqref{obs:harmonic-weight}, we obtain that
\begin{equation}\label{obs:screened-upper}
\Phi_r\leq Cr^{-3}h\text{ \ in \ }A_r\text{ \ for \ }0<r\leq D_*<1.
\end{equation}
Applying Lemma~\ref{lem:3.2} with $I=F$ and $u=r$, we obtain from \eqref{eq:intcom} that
\begin{equation}\label{obs:small-exterior}
\int_{A_r}\rho dx\leq Cr^{-3},\ \  \int_{A_r}\rho^{5/3} dx\leq Cr^{-7}\text{ \ for \ }0<r\leq D_*<1.
\end{equation}

Set
\begin{equation}\label{obs:fixed-radius}
r_0:=D_*/2\in(0,1).
\end{equation}
It follows from \eqref{obs:small-exterior} and \eqref{obs:fixed-radius} that for every $r\geq r_0$,
\begin{equation*}
\begin{aligned}
\int_{A_r}\rho dx\leq&\int_{A_{r_0}}\rho dx\leq Cr_0^{-3}\leq C(1+r^{-3}),\\
\int_{A_r}\rho^{5/3} dx\leq&\int_{A_{r_0}}\rho^{5/3} dx\leq Cr_0^{-7}\leq C(1+r^{-7}).
\end{aligned}
\end{equation*}
Combining this estimate with \eqref{obs:small-exterior}, we obtain that
\begin{equation}\label{obs:exterior}
\int_{A_r}\rho dx\leq C(1+r^{-3}),\ \  \int_{A_r}\rho^{5/3} dx\leq C(1+r^{-7})\text{ \ for \  }r>0.
\end{equation}

By the maximum principle, we derive from \eqref{obs:small-screening} that $\|\Phi_{r_0}-\Phi_{r_0}^{\TF}\|_{L^\infty(A_{r_0})}\leq C$. Moreover, we deduce from \eqref{eq:screen1} that for $r>r_0$,
\begin{equation}\label{obs:large-radius-decomposition}
\begin{aligned}
\Phi_r-\Phi_r^{\TF}=&\Phi_{r_0}-\Phi_{r_0}^{\TF}-\bigl[\ind_{\Omega_r\setminus\Omega_{r_0}}(\rho-\rho_{\TF})\bigr]*|\cdot|^{-1}\text{ \ in \  }A_{r_0}
\end{aligned}
\end{equation}
Since $|\ind_{\Omega_r\setminus\Omega_{r_0}}(\rho-\rho_{\TF})|\leq\ind_{A_{r_0}}(\rho+\rho_{\TF})$ in $\R^3$ and $\partial\Omega_r\subset\overline{A_{r_0}}$ for $r\geq r_0$, we obtain from \eqref{obs:large-radius-decomposition} that for $r\geq r_0$,
\begin{equation*}
\sup_{\partial\Omega_r}|\Phi_r-\Phi_r^{\TF}|\leq C+\|(\ind_{A_{r_0}}(\rho+\rho_{\TF}))*\coul\|_\infty\leq C,
\end{equation*}
where we have used \eqref{obs:coulomb-interpolation}, \eqref{obs:tf-upper-tail} and \eqref{obs:exterior} in the last inequality. Together with \eqref{obs:small-screening}, this proves \eqref{eq:screebound}.

Since $N\geq Z$, we obtain from \eqref{obs:core-mass} and Theorem~\ref{thm:1.1} that for $0<r\leq r_0$,
\begin{equation*}
\begin{aligned}
\Big|\int_{A_r}(\rho-\rho_{\TF})dx\Big|=&\Big|N-Z-\int_{\Omega_r}(\rho-\rho_{\TF})dx\Big| \leq(N-Z)+\left|\int_{\Omega_r}(\rho-\rho_{\TF}) dx\right|\\
\leq&C(1+r^{-3+10^{-5}}).
\end{aligned}
\end{equation*}
It follows from \eqref{obs:tf-upper-tail} and \eqref{obs:exterior} that for $r\geq r_0$,
\begin{equation*}
\Big|\int_{A_r}(\rho-\rho_{\TF}) dx\Big|\leq\int_{A_{r_0}}(\rho+\rho_{\TF}) dx\leq C.
\end{equation*}
This proves \eqref{eq:dmass} and completes the proof of Theorem~\ref{thm:screen}.
\end{proof}

\subsection{Proof of Theorems~\ref{thm:radius} and \ref{thmm:ener}}
This subsection is devoted to proving  Theorems~\ref{thm:radius} and \ref{thmm:ener}. We first estimate the mass of $\rho_{\TF}$ in $A_s\setminus A_t$ for $0<s<t$.

\begin{lemma}\label{obs:tf-slope}
Let $M:=\max_{1\leq j\leq K}Z_j$. For every $0<s\leq t$ with $Ms^3\geq1$, we have
\begin{equation}\label{obs:tf-shell-lower}
\int_{A_s\setminus A_t}\rho_{\TF} dx\geq c_K(t-s)t^{-4}.
\end{equation}
In particular, for every $r>0$ with $Mr^3\geq1$, we have
\begin{equation}\label{obs:tf-two-sided}
c_Kr^{-3}\leq \int_{A_r}\rho_{\TF} dx\leq C_Kr^{-3}.
\end{equation}
\end{lemma}

\begin{proof}
Fix $0<s<t$ with $Ms^3\geq1$ and choose $i_0$ such that $Z_{i_0}=M$. For $\omega\in\mathbb S^2$, consider the ray $R_{i_0}+u\omega$ for $u\geq0$. Each ball $\overline{B(R_j,t)}$ intersects this ray in a set of length at most $2t$. Since the $K$ balls cover a total length of at most $2Kt$ on each ray, for every $\omega\in\mathbb S^2$ there exists $u=u(\omega)\leq(2K+1)t$ with $u\geq0$ such that $R_{i_0}+u\omega\notin\Omega_t$. Since $d(R_{i_0})=0$, we may take $u_t$ to be the first $u$ with $d(R_{i_0}+u\omega)=t$ and $u_s$ to be the last $u<u_t$ with $d(R_{i_0}+u\omega)=s$. By the $1$-Lipschitz continuity of $d$, we have
\begin{equation*}
u_t-u_s\geq t-s,\qquad s<d(R_{i_0}+u\omega)<t\text{ \ for \ }u_s<u<u_t.
\end{equation*}
For $u_s(\omega)<u<u_t(\omega)$, we have $s<d(R_{i_0}+u\omega)\leq u\leq(2K+1)t$, and hence $Mu^3\geq Ms^3\geq1$. It follows from \eqref{eq:tfe}, \eqref{eq:acomparison} and \eqref{eq:aestimates} that $\rho_{\TF}(R_{i_0}+u\omega)u^2\geq\rho_{i_0}^{\rm at}(R_{i_0}+u\omega)u^2\geq cu^{-4}\geq c_Kt^{-4}$. We then obtain that
\begin{equation*}
\int_{A_s\setminus A_t}\rho_{\TF} dx
=\int_{\mathbb S^2}\int_{\{u\geq0: s<d(R_{i_0}+u\omega)\leq t\}}\rho_{\TF}(R_{i_0}+u\omega)u^2 du d\omega\geq c_K(t-s)t^{-4}.
\end{equation*}
This proves \eqref{obs:tf-shell-lower} for $s<t$, and equality holds for $s=t$. Taking $s=r$ and $t=2r$ in \eqref{obs:tf-shell-lower}, we have $\int_{A_r}\rho_{\TF} dx\geq\int_{A_r\setminus A_{2r}}\rho_{\TF} dx\geq(c_K/16)r^{-3}$ for $Mr^3\geq1$. Together with \eqref{obs:tf-upper-tail}, this completes the proof of Lemma~\ref{obs:tf-slope}.
\end{proof}

Combining the mass comparison in Theorem~\ref{thm:screen} with Lemma~\ref{obs:tf-slope}, we now prove Theorem~\ref{thm:radius} on the radius comparison.

\begin{proof}[Proof of Theorem~\ref{thm:radius}]
Let $c_K$ be as in Lemma~\ref{obs:tf-slope}, and let $C_K\geq\max\{1,c_K\}$ be sufficiently large so that \eqref{eq:dmass}, \eqref{obs:tf-upper-tail} and \eqref{obs:exterior} hold. For $\kappa\geq4C_K$, we obtain from \eqref{obs:tf-upper-tail} and \eqref{obs:exterior} that
\begin{equation}\label{obs:radius-upper-masses}
\begin{aligned}
\int_{A_{(4C_K/\kappa)^{1/3}}}\rho dx\leq C_K+\kappa/4\leq\kappa/2,\ \ \int_{A_{({4C_K}/{\kappa})^{1/3}}}\rho_{\TF} dx\leq\kappa/2.
\end{aligned}
\end{equation}

Take sufficiently large $A_K>1$ so that $A_Kc_K\geq4K$. Since $M\geq Z/K\geq A_K\kappa/K$ by the conditions of Theorem~\ref{thm:radius}, we have $Mc_K/(4\kappa)\geq1$. This then gives from \eqref{obs:tf-two-sided} that
\begin{equation}\label{obs:radius-lower-tf-mass}
\int_{A_{\left(\frac{c_K}{4\kappa}\right)^{1/3}}}\rho_{\TF} dx\geq4\kappa.
\end{equation}
Moreover, it follows from \eqref{eq:dmass} that there exists a sufficiently large $\kappa_0\geq\max\{1,4C_K\}$, depending only on $K>0$, such that for every $\kappa\geq\kappa_0$,
\begin{equation*}
\Big|\int_{A_{(c_K/(4\kappa))^{1/3}}}(\rho-\rho_{\TF}) dx\Big|\leq C_K\Big(1+\left(\frac{c_K}{4}\right)^{-1+10^{-5}/3}\kappa^{1-10^{-5}/3}\Big) \leq \kappa .
\end{equation*}
Together with \eqref{obs:radius-lower-tf-mass}, this implies that
\begin{equation}\label{obs:radius-lower-rho-mass}
\int_{A_{(c_K/(4\kappa))^{1/3}}}\rho dx\geq 4\kappa-\kappa\geq3\kappa\text{ \ for \  }\kappa\geq\kappa_0.
\end{equation}
Since $r\mapsto\int_{A_r}\rho dx$ and $r\mapsto\int_{A_r}\rho_{\TF} dx$ are continuous and nonincreasing, we obtain from \eqref{eq:Rf}, \eqref{obs:radius-upper-masses}--\eqref{obs:radius-lower-rho-mass} that
\begin{equation}\label{obs:radius-bracket}
R_\rho(\kappa),R_{\TF}(\kappa)\in\Big[\left(\frac{c_K}{4\kappa}\right)^{1/3},\left(\frac{4C_K}{\kappa}\right)^{1/3}\Big]\text{ \ for \  }\kappa\geq\kappa_0,
\end{equation}
which proves \eqref{eq:radiusb}.

Set $s:=R_\rho(\kappa)>0$ and $t:=R_{\TF}(\kappa)>0$. It then follows from \eqref{eq:Rf} that $\int_{A_s}\rho dx=\int_{A_t}\rho_{\TF} dx=\kappa$. Since  $Mc_K/(4\kappa)\geq1$, we obtain from \eqref{obs:radius-bracket} that $M\min\{s,t\}^3\geq1$. Applying \eqref{obs:tf-shell-lower} between $s$ and $t$, we thus deduce from \eqref{eq:dmass} and \eqref{obs:radius-bracket} that
\begin{equation*}
\begin{aligned}
c_K\left(\frac{4C_K}{\kappa}\right)^{-4/3}|s-t|
&\leq c_K\max\{s,t\}^{-4}|s-t|\leq\Big|\int_{A_s}\rho_{\TF} dx-\int_{A_t}\rho_{\TF} dx\Big|\\
&=\Big|\int_{A_s}(\rho_{\TF}-\rho) dx\Big|\leq C_K(1+s^{-3+10^{-5}})\\
&\leq C_K\Big[1+\left(\frac{c_K}{4}\right)^{-1+10^{-5}/3}\Big]\kappa^{1-10^{-5}/3}\text{ \ for \ }\kappa\geq\kappa_0\geq1.
\end{aligned}
\end{equation*}
This implies that
\begin{equation*}
|R_\rho(\kappa)-R_{\TF}(\kappa)|\leq C_K\kappa^{-4/3}\kappa^{1-10^{-5}/3}=C_K\kappa^{-1/3-10^{-5}/3}\text{ \ for \  }\kappa\geq\kappa_0.
\end{equation*}
We thus complete the proof of Theorem~\ref{thm:radius}.
\end{proof}

We use the lower bound on $R_\rho$ in Theorem~\ref{thm:radius} to choose the cutoff radius in the proof of Theorem~\ref{thmm:ener}.

\begin{proof}[Proof of Theorem~\ref{thmm:ener}]
Fix $0<\kappa\leq N$. Let $c_K$ be the constant in the lower bound \eqref{eq:radiusb}. Increase $\kappa_0$ in Theorem~\ref{thm:radius} if necessary so that $(c_K/2)\kappa_0^{-1/3}\leq D_*$, where $D_*$ is as in Lemma~\ref{iter:finite-screening}. Set $m:=\max\{\kappa,\kappa_0\}$.

Suppose first that $Z\geq A_Km$. Since $m\geq\kappa_0$ and $N\geq Z\geq A_Km>m$, by Theorem~\ref{thm:radius} we have $R_\rho(m)\geq c_Km^{-1/3}$.  Since $r\mapsto\int_{A_r}\rho dx$ is continuous, we obtain from \eqref{eq:Rf} that
\begin{equation}\label{ener:selected-radius}
r_*:=\tfrac12c_Km^{-1/3}\leq D_*,\ \ 
\int_{A_{2r_*}}\rho dx\geq\int_{A_{R_\rho(m)}}\rho dx=m,
\end{equation}
where we have used $2r_*\leq R_\rho(m)$ and $A_{R_\rho(m)}\subset A_{2r_*}$ in  the mass inequality.
For $r>0$, we choose $\chi_r,\eta_r\in C^\infty(\R^3;[0,1])$ such that
\begin{equation}\label{ener:cutoff-definition}
\chi_r^2+\eta_r^2\equiv1\text{ \ in \  }\R^3,\ \  \chi_r\equiv1\text{ \ in \  }\Omega_r,\ \  \chi_r\equiv0\text{ \ in \  }A_{2r},
\end{equation}
and
\begin{equation}\label{ener:cutoff-gradient}
|\nabla\chi_r|^2+|\nabla\eta_r|^2\leq Cr^{-2}\ind_{A_r\setminus A_{2r}}\text{ \ in \  }\R^3.
\end{equation}

Since $\ind_{A_{2r_*}}\leq\eta_{r_*}^2\leq\ind_{A_{r_*}}$ in $\R^3$, we deduce from \eqref{obs:small-exterior} and \eqref{ener:selected-radius} that
\begin{equation}\label{ener:removed-mass}
\kappa\leq m\leq\int_{A_{2r_*}}\rho dx\leq\int_{\R^3}\eta_{r_*}^2\rho dx\leq\int_{A_{r_*}}\rho dx\leq Cr_*^{-3}\leq Cm.
\end{equation}
Following the argument in \cite[proof of Theorem~3.8]{SolovejHF}, we use $\chi_{r_*}^2\rho$ as a trial density. Since $\int_{\R^3}\chi_{r_*}^2\rho dx=N-\int_{\R^3}\eta_{r_*}^2\rho dx\leq N-\kappa$ by \eqref{ener:cutoff-definition} and \eqref{ener:removed-mass}, we obtain from \eqref{eq:energym} and \eqref{eq:inf} that
\begin{equation}\label{ener:cutoff-competitor}
I_V(N-\kappa)\leq I_V\left(\int_{\R^3}\chi_{r_*}^2\rho dx\right)\leq\mathcal E_V(\chi_{r_*}^2\rho).
\end{equation}
From \eqref{obs:harmonic-weight}, \eqref{obs:screened-upper} and \eqref{ener:selected-radius}, we have $\Phi_{r_*}\leq Cr_*^{-4}$ in $A_{r_*}$. Since $c^{\TF}z^{5/3}-c^{\mathrm D}z^{4/3}\geq-Cz$ for $z\geq0$, we also have $\mathcal E_0(\eta_{r_*}^2\rho)\geq-C\int_{\R^3}\eta_{r_*}^2\rho dx$. Applying \eqref{eq:binding} with $I=\{1,\ldots,K\}$, $r=r_*$ and the cutoffs in \eqref{ener:cutoff-definition}, we deduce from \eqref{ener:cutoff-gradient} and \eqref{ener:cutoff-competitor} that for $m\geq1$,
\begin{equation}\label{ener:large-charge-removal}
\begin{aligned}
I_V(N-\kappa)-I_V(N)
&\leq\mathcal E_V(\chi_{r_*}^2\rho)-\mathcal E_V(\rho)\\
&\leq C(1+r_*^{-2})\int_{A_{r_*}\setminus A_{2r_*}}\rho dx+C(1+r_*^{-4})\int_{\R^3}\eta_{r_*}^2\rho dx\\
&\leq C(1+r_*^{-2})r_*^{-3}+C(1+r_*^{-4})m\leq Cm^{7/3},
\end{aligned}
\end{equation}
where we have used \eqref{ener:removed-mass} in the third inequality and \eqref{ener:selected-radius} together with $m\geq1$ in the last inequality.

If $Z<A_Km$, then Theorem~\ref{thm:1.1} gives that $N\leq Z+C_K<A_Km+C_K$. Since $I_V(N-\kappa)\leq I_V(0)=0$ by \eqref{eq:energym}, we deduce from \eqref{eq:2.44} that
\begin{equation}\label{ener:small-charge-removal}
I_V(N-\kappa)-I_V(N)\leq-I_V(N)\leq C(Z^{7/3}+N)\leq C_Km^{7/3}.
\end{equation}
Since $m^{7/3}\leq\kappa^{7/3}+\kappa_0^{7/3}$ and $\kappa_0$ depends only on $K$, the upper bound in \eqref{eq:ener1} follows from \eqref{ener:large-charge-removal} and \eqref{ener:small-charge-removal}. Together with \eqref{eq:energym}, this completes the proof of Theorem~\ref{thmm:ener}.
\end{proof}

\section*{Acknowledgment}
The author thanks Professor Yujin Guo for helpful discussions on this work. He also thanks Professor Phan Th\`{a}nh Nam, whose online talk drew the author's attention to this problem.

\section*{Data Availability}
No data were used for the research described in the article.

\section*{Declaration of AI-assisted tools}
This paper was written by the author and was not generated by artificial intelligence. During its preparation, the author used AI tools as auxiliary aids for exploratory discussions, preliminary proof drafting, and language editing. The author takes full responsibility for the mathematical validation, the final proof choices, and the final text.

\end{document}